%% file: main.tex
\documentclass[manuscript, longbib]{aastex701}

\usepackage{amsmath}
\usepackage{comment}
\shorttitle{Crater Scaling Relationships for Gas Impingement}
\shortauthors{Ballouz et al.}
\graphicspath{{./}{Figures/}}

\begin{document}
\title{Crater Scaling Relationships for Gas Impingement on Planetary Regolith}
\author[orcid=0000-0002-1772-1934, gname=Ronald, sname=Ballouz]{Ronald L. Ballouz}
\affiliation{Johns Hopkins Applied Physics Laboratory,  Laurel, MD 20723, USA}
\email[show]{ronald.ballouz@jhuapl.edu}

\author[orcid=0000-0002-3578-7750, gname=Olivier S., sname=Barnouin]{Olivier S. Barnouin}
\affiliation{Johns Hopkins Applied Physics Laboratory,  Laurel, MD 20723, USA} 
\email{Olivier.Barnouin@jhuapl.edu}

\author[gname=Mitsuhisa, sname=Baba]{Mitsuhisa Baba}
\affiliation{JAXA, Sagamihara, Kanagawa , 252-5210, Japan}
\email{}

\author[gname=Vicki, sname=Toy-Edens]{Vicki L. Toy-Edens}
\affiliation{Johns Hopkins Applied Physics Laboratory,  Laurel, MD 20723, USA}
\email{vicki.toy-edens@jhuapl.edu}

\author[gname=Noble, sname=Hatten]{Noble Hatten}
\affiliation{Johns Hopkins Applied Physics Laboratory,  Laurel, MD 20723, USA}
\email{noble.hatten@jhuapl.edu}

\author[orcid=0000-0001-8369-8587, gname=Parvathy, sname=Prem]{Parvathy Prem}
\affiliation{Johns Hopkins Applied Physics Laboratory,  Laurel, MD 20723, USA} 
\email{Parvathy.Prem@jhuapl.edu}

\author[orcid=0009-0004-7820-107X, gname=Maarij, sname=Qureshi]{Maarij Qureshi}
\affiliation{Johns Hopkins Applied Physics Laboratory,  Laurel, MD 20723, USA}
\email{maarijq1@terpmail.umd.edu}

\author[orcid=0000-0002-7538-7903, gname=Cecily, sname=Sunday]{Cecily Sunday}
\affiliation{Institut Supérieur de l'Aéronautique et de l'Espace, Université de Toulouse, Toulouse, 31055, France}
\email{cecily.sunday@isae-supaero.fr}

\begin{abstract}
Plume-surface interaction (PSI) effects remains a major uncertainty for operations on regolith-covered planetary bodies. Although spacecraft exhaust can erode surfaces, form craters, and accelerate ejecta, existing PSI relationships are largely derived for scenario-specific lunar and Martian landing conditions and are not yet reliably generalizable. Resolving this problem is important not only for spacecraft safety, but also because controlled thruster interactions could provide an active geotechnical experiment during asteroid reconnaissance. Measurements of thruster-formed craters and ejecta may constrain regolith density, strength, porosity, and particle properties that are otherwise difficult to determine remotely. Such information is particularly relevant to planetary defense, where uncertainty in near-surface properties can strongly affect the outcome of kinetic-impact mitigation and the interpretation of momentum transfer. We investigate gas-driven excavation using near-vacuum, pulsed-N$_2$ experiments in a cohesionless asteroid regolith analog and comparisons with a low-gravity drop-tower experiment and historical chemical-thruster tests. We adapt projectile-cratering $\pi$-scaling, and find that these relationships can describe gas crater volume as a function of gas, target, gravity, and nozzle properties. A common gravity-regime relationship describes the laboratory data and is consistent with an experimental low-gravity result. Generalizing the $\pi$-scaling relationship for cohesive targets shows consistency with Apollo-era chemical-thruster cases. These results provide a framework for using spacecraft thrusters as active geotechnical tools during small-body reconnaissance.
\end{abstract}


\input{_01_introduction}
\input{_02_methods}

\input{_03_results}
\input{_04_discussion_conclusions}

\input{_05_acknowledgements}

\bibliography{_06_references}
\bibliographystyle{aasjournalv7}

\input{_07_appendix}


\end{document}

%% file: _01_introduction.tex
\section{Introduction} 
\label{sec:introduction}

The surfaces of rocky and icy small solar system bodies can be covered by a layer of unconsolidated pebbles, cobbles, and boulders, that is often termed regolith. The geotechnical properties of a small body's regolith and interior, such as particle size distribution and shape, bulk density, porosity, cohesion, and angle of internal friction, are critical aspects for the design of \textit{in-situ } exploration \citep{Lorenz2011}  and planetary defense missions \citep{Stickle2022}. Historically, little is known about an asteroid or comet's regolith geotechnical properties \textit{a-priori} and extrapolations from prior surface exploration missions are typically used to justify design choices and operation strategies. This has often led to surprising outcomes, even for targets previously thought to be well characterized \citep{Lauretta+2019, lauretta_spacecraft_2022, walsh_near-zero_2022, Arakawa_2020, Biele2015}. 

In the event of the discovery of a hazardous asteroid, spacecraft reconnaissance can provide valuable information on the asteroid’s physical properties in order to formulate an optimized mitigation strategy.  The near-surface and interior geotechnical properties of an asteroid are critical to the design of planetary defense strategies. Asteroid mitigation strategies typically fall into four general categories \citep{nitep2021report}: civil defense, slow-push/pull techniques, kinetic impact, and nuclear detonation. Depending on the nature and size of the incoming object as well as the warning time before impact, one of these different approaches would become preferred \citep{OWL2023}. Recently, the Double Asteroid Redirect Test (DART) mission demonstrated the kinetic impact mitigation technique by impacting the binary companion of the rubble-pile asteroid Didymos, and measured the momentum enhancement factor ($\beta$) from the impact \citep{Daly+2023}. 

Numerical simulations of kinetic impactors have shown that an asteroid’s surface and interior geotechnical properties, such as porosity, cohesion, and angle of internal friction in the near-surface have the largest effect on $\beta$ and deflection velocity \citep{Walker2011, Syal2016, Raducan2022RubblePile, Owen2022, Luther2022}.  Prior to the DART demonstration, the mission's impact working group found that massive uncertainties in near-surface geotechnical properties drive a factor of five uncertainty in momentum enhancement from DART \citep{Stickle2022}. For kinetic impact mitigation, material properties and geology within several meters of the impact site dominate over internal structure \citep{Stickle2017, Raducan2022RubblePile, Ormo2022}. For nuclear mitigation, numerical studies have found that the interior structure of the asteroid target plays a role in controlling the outcome of a deflection attempt \citep{King2021}.

While some geotechnical properties may be determined from remote sensing, such as boulder size and shape properties, others can only be poorly constrained, such as sub-surface porosity and strength.  For small bodies, variations in porosity and regolith strength have lead to a diversity of outcomes during spacecraft interactions with surfaces during landing, roving, or sampling. For example, the Hayabusa spacecraft experienced meter-high bouncing upon touching down on the surface of the asteroid Itokawa \citep{Yano2006}, the Philae lander experienced  displacements of hundreds of meters upon contaciting the surface of comet 67P/CG \citep{Biele2015}, and  the OSIRIS-REx spacecraft experienced nearly no resistance as it penetrated into the surface of the asteroid Bennu \citep{walsh_near-zero_2022}.  

These examples highlight how spacecraft-surface interactions have provided serendipitous, yet valuable, insight into the surface geoechnical properties of small bodies. However, to meet the demands of a robust planetary defense strategy \citep{NASA_PD_Plan, US_National_Plan} requires the development of technologies that can determine surface geotechnical properties of Potentially Hazardous Asteroids (PHAs) through spacecraft rendezvous reconnaissance. Recent exploration of the near-Earth asteroids (NEAs) Bennu and Ryugu by the OSIRIS-REx and Hayabusa2 missions, respectively, highlight a potential pathway to characterize the surface and interior geotechnical properties of asteroids through spacecraft rendezvous missions. 

\subsection{Cratering as a probe of small body regolith geotechnical properties}

Impact craters are ubiquitous on planetary surfaces. Decades of laboratory and numerical experiments have demonstrated that the physical features of a crater, such as its size, depth, and shape, can be determined based on the properties of the impactor  (e.g., size, speed, density) and target (mineralogy and geotechnical properties) \citep{Holsapple+1987, Housen+2018}.  For asteroids, crater-scaling relationships can relate a crater’s physical properties to the asteroid’s surface gravity, density, and strength and the impactor’s radius, density, and impact speed. When the impactor and impact parameters are known, then an observed crater’s size may be used to estimate target properties. 

JAXA's Hayabusa2 mission conducted the Small Carry-on Impactor (SCI) experiment to create an artificial impact crater to expose fresh, subsurface material for sample collection and to investigate the cratering process on a small asteroid \citep{Arakawa+2017}.  The experiment launched a 2-kg copper projectile at $\sim$2 km/s into Ryugu's surface. The Deployable Camera 3 (DCAM3) \citep{Ogawa+2017} witnessed the impact and subsequent formation of the SCI-crater \citep{arakawa_artificial_2020}. Before-and-after image comparisons of the impact site using Optical Navigation Camera (ONC) data revealed a $\sim$ 15 m diameter crater; when combined with DCAM3 observations, these data revealed the low strength of the Ryugu regolith at the point of impact  \citep{arakawa_artificial_2020}. By using laboratory-calibrated crater scaling relationships, \citep{arakawa_artificial_2020} used SCI crater measurements to determine that the regolith of Ryugu was had a cohesive strength of only 1.3 Pa, and crater formation was likely controlled by the gravity of the asteroid, despite the gravity only being $\sim0.11$ mm/s$^2$ \citep{Watanabe+2019}. 

NASA's OSIRIS-REx mission sampled the surface of asteroid Bennu through a touch-and-go (TAG) maneuver that involved: 1) contact of the TAG sample acquisition mechanism (TAGSAM), which articulated from the main spacecraft body via a robotic arm, 2) relatively un-impeded penetration of the TAGSAM into the Bennu regolith for $\sim$1 s, 3) the firing of pressurized nitrogen gas (N$_2$) from the TAGSAM to entrain regolith in the gas which circulated back into the sample collection chamber, and 4) the firing of four attitude control system 4.5-N thrusters on the main spacecraft body to depart from the surface \citep{lauretta_spacecraft_2022, walsh_near-zero_2022, Bierhaus2026}. The combination of TAGSAM penetration, TAGSAM gas firing, and thruster firing led to the formation of an elliptical crater with dimensions of $9.0$ by $6.5$ m, with its long axis oriented from the asteroid's north-to-south poles. The average diameter of the TAG crater was $7.8\pm1.8$ m, its depth was $0.68\pm0.1$ m., and the the total displaced volume was $12.2\pm0.9$ m$^3$. \citet{lauretta_spacecraft_2022} proposed that most of the excavated volume was due to the firing of the TAGSAM's N$_2$, and used crater scaling relationships developed for projectiles to estimate that the near-surface bulk density of the regolith at the sampling site was $0.5-0.7$ g/cm$^3$. This finding is consistent with independent estimates of the bulk density of the upper 10-cm of the touchdown site derived from analysis of the peak force registered by the spacecraft's intertial measurement units (IMU's), which suggested a bulk density of $0.58-0.87$ g/cm$^3$ \citep{walsh_near-zero_2022,Ballouz_2021} and nearly cohesionless surface. 

Though, more recent analysis by \citet{Bierhaus2026} demonstrated that TAGSAM initially formed a crater smaller than $\sim1$ m in radius before thruster activation deposited four regions of high-pressure gas 1-6 meters from the contact location. As a result, the TAGSAM crater expanded into a region undergoing active erosion from the thrusters, and its final dimensions were increased by thruster effects. The total erosion and redeposition was influenced by the pre-existing mass distribution and topography of the area. Furtherore, \citet{Bierhaus2026} concluded that variability in surface response to the thrusters indicated that the material properties of the sampling site regolith varied laterally.

\subsection{Cratering as a probe of small body interior geotechnical properties}
The irregular mass distribution inside rubble-pile asteroids is manifested in their gravity field. Therefore, gravity field information can be exploited to deduce their internal structure with a technique called gravity inversion (GI). The key advantage of GI is that gravity measurements can be obtained remotely and without any special instrument onboard, from the natural motion of the visiting spacecraft (orbit determination).  However, a spacecraft stationed afar can only resolve this gravity field in low resolution due to the weak signal from the asteroid's microgravity environment. A spacecraft would need to be in an orbit with a semi-major axis of $\sim$1 body-diameter to be able to measure the spherical expansion terms of the body's gravity to degree and order 3. Through the serendipitous discovery of sporadic particle ejection events on the asteroid, the OSIRIS-REx mission was also able to better constrain the interior density distribution of the asteroid by tracking particles \citep{Lauretta2019sciencea}. The episodes of particle ejections enabled motion tracking of $\sim$300 particles of a few mm-cm in size to extract an \textit{a-priori} unanticipated, high-resolution gravity field \citep{Chesley20202019je00, Scheeres2020}, revealing that Bennu's interior mass is heterogeneously distributed. In addition, ejected particle photometry and dynamical modeling enabled the measurement of  particles shape properties and the Bennu surface’s coefficient of restitution, respectively \citep{Chesley20202019je00}.  

These activities demonstrate the utility of tracking ejected particles for characterizing asteroid physical properties. However, it is unclear if the sporadic natural phenomenon observed at Bennu is common across all asteroid sizes and types, or if natural particle ejections are specific to C-complex asteroids such as Bennu. Multiple hypotheses were presented to explain particle ejections, such as micrometeoroid impacts, thermal fatigue, and de-volatilization \citep{Bottke20202019je00, Molaro20202019je00, Lauretta2019sciencea}.  These mechanisms are dependent on asteroid physical and orbital properties; therefore, it is unclear whether tracking if a reconnaissance mission to a PHA can rely on such phenomena to characterize the interior of the asteroid. An alternative approach would be to utilize cratering to produce artificially ejected particles that can similarly be tracked for asteroid particle, surface, and interior characterization. 

Here, we advocate for characterizing a small body's surface and interior geotechnical properties through thruster-based cratering. As thrusters are a necessary component to any spacecraft exploring small bodies, thruster-based regolith excavation could be utilized to form a crater and generate trackable ejecta at multiple points on the surface of a PHA. Characterization of the thruster-formed crater and tracking of ejected particles can be used to estimate the geotechnical properties of a PHA.  

\subsection{Spacecraft Thruster Plume Surface Interaction}

Spacecraft thruster plume surface interaction (PSI) has been studied in the context of lunar landers such as those used by Apollo missions, and, more recently, landers from NASA's Commercial Lunar Payload Service (CLPS) initiative, ISRO, and CNSA \citep{Morris2015, CLPS,kim2025survey}, as the ejection of dust can create a challenging environment for spacecraft landings. These missions have helped us understand the mechanisms driving dust erosion during lunar landings. In conjunction with theoretical and experimental works \citep[see exhaustive reviews in][]{kim2025survey,cuesta2025plume}, these operations have revealed distinct physical mechanisms by which spacecraft thrusters mobilize regolith:
\begin{enumerate} 
\item Viscous erosion (VE) occurs when gas flowing parallel to or obliquely across a surface exerts shear stresses that exceed gravitational, inter-particle friction, and  inter-particle cohesive forces \citep{Shorthill1976,Christensen1967}.  VE results in the stripping of the top-most layer of fines as they are accelerated to high speeds away from the spacecraft. This is the main process that generates dust seen in images of lunar landings \citep[e.g.,][]{Sunday2025}. 
\item Bearing capacity failure (BCF) occurs when the downward pressure of gas exceeds the regolith's bearing capacity. 
\item Diffusion driven flow (DDF) occurs when the gas penetrates regolith pore space, leading to the build-up of pressure in the sub-surface.  
In conjunction, BCF and DDF results in the formation of craters when gas pressure exceeds the shear strength of the soil \citep{Metzger2009}. This effect is localized to the region in proximity to where the gas impinges on the surface, as opposed to the near- and far-field effects of VE. 
\item Diffused gas eruption (DGE) occurs when gases rapidly escape from regolith pore space, which occurs when a large pressure gradient develops when the thrusters are shut-off. This leads to the ejection of regolith particles \citep{Metzger2009}. 
\item Diffusive gas explosive erosion (DGEE) or ``explosive erosion'' dominates when supersonic pulsed jets impinges the regolith, generating high pore pressure gradient forces that exceeds the weight of the soil \citep{Mehta2011}. DGEE played a significant role in the alteration of the landing site of the Phoenix lander when it descended on the Martian polar region \citep{Mehta2011}. 
\end{enumerate}

For the Lunar- and Martian- landing problems, \citet{rahimi2020} identified that viscous erosion was the dominant mechanism that controlled PSI, as the shear stress, $\tau$ from the thruster plume can exceed soil shear strength, $\tau_C$, leading to aerodynamic entrainment of dust. Works by \citet{Metzger2009, Metzger2024a, Metzger2024b} extended foundational treatments of viscous erosion scaling by \citet{roberts1963}, which formulates the mass erosion rate, $\dot{m}$, as:
\begin{equation}
    \dot{m} = 2 \frac{\tau - \tau_C}{b u},  
    \label{eq:roberts}
\end{equation}
where $b$ is the fraction of the velocity that particles gain from the gas phase, $u$ is the gas phase velocity. $\tau_C$ can be defined through the Mohr-Coulomb criterion, 
\begin{equation}
    \tau_C = C + P\tan\phi,
    \label{eq:mohr_coulomb}
\end{equation}
where $C$ is the cohesive strength of the soil, $P$ is the normal stress on the regolith which would be a function of the gas static pressure and gravity, and $\phi$ is the angle of friction of the regolith. The coefficient, $b$, is a function of thruster gas and engine properties, hover altitude, and average regolith particle density and size \citep[see Eqs.~7\&8 of][]{rahimi2020}. Recent studies of lunar landing PSI have used Eq.~\ref{eq:roberts} in conjunction with hybrid computational fluid dynamics (CFD) and direct simulation Monte Carlo (DSMC) techniques to model thruster plume expansion and the trajectories of ejected dust particles from viscous erosion \citep{Morris2015,rahimi2020,Sunday2025}. 

Recently, \citet{Metzger2024a,Metzger2024b} demonstrated that the viscous erosion rate is related to the ratio of gas kinetic energy to the combined potential and cohesive energy of the surface, deriving a formulation for the volumetric erosion rate, $\dot{V}$ as a function of thruster and surface properties. The formulation was further refined by \citet{Dotson2025a}, who systematically studied the dependence of $\dot{V}$ on surface geotechnical properties and derived the following relationship for situations where regolith $C < 1000$ Pa: 
\begin{equation}
    \dot{V} \approx 2\pi \frac{\delta U^2 A}{0.5  \rho g D + C}.
    \label{eq:dotson_v_dot}
\end{equation}
$\delta$ is gas jet density, $U$ is the gas jet velocity, $A$ is the nozzle exit area, $\rho$ is the regolith bulk density, $g$ is the local gravitational acceleration, and $D$ is the regolith mean diameter. Further experimental work by \citet{Dotson2025b} investigated the influence of ambient pressure outside the nozzle, $P_a$, on PSI, as gaseous flows have different behaviors at reduced ambient pressures due to gas molecule interactions. The degree to which a fluid can be treated as a continuum is typically parameterized through the Knudsen number, $\mathrm{Kn}$, a dimensionless number that is the ratio of the molecular mean free path, $\lambda$, to a characteristic length of gas flow, $L$:
\begin{equation}
    \mathrm{Kn} = \frac{\lambda}{L} = \frac{k_B T_a}{\sqrt{2}\pi d_a^2 P_a D},
\end{equation}
where $k_B$ is the Boltzmann constant, $T_a$ is the ambient temperature, and $d_a$ is the diameter of the atmospheric particle \citep{Mehta2011,capecelatro_wagner2024,Dotson2025b}. In the continuum regime ($\mathrm{Kn} \lesssim 0.1$), gas flows behave like a continuous fluid where collisions between molecules are frequent. In the free-molecular flow regime ($\mathrm{Kn} \gtrsim 10$), the mean free path of a gas molecule is much larger than the characteristic length, so flow behavior is dominated by individual molecular motion, leading to the breakdown of continuum assumptions. For $0.1 \lesssim Kn \lesssim 10$, the mean free path of a gas molecule is comparable to the characteristic length scale, leading to complex interactions between molecular and continuum effects. By varying $P_a$,  \citet{Dotson2025b} showed that increasing  $\mathrm{Kn}$ leads to larger $\dot{V}$ with a maximum value occurring in the transition region, and provided an updated formulation for $\dot{V}$ that included a functional dependence $f(\mathrm{Kn})$. 

The importance of $P_a$ in influencing the outcome of PSI was also explored by \citet{Mehta2011} in the context of  understanding the mechanics of PSI that led to the excavation of sub-surface polar ice on Mars. \citet{Mehta2011} experimentally recreated the conditions of NASA's Phoenix lander's descent to the surface of the Mars polar regions in the laboratory by ensuring dynamic similitude across multiple dimensionless constants that took into account Phoenix's thruster configuration and the Martian surface and atmospheric conditions.  \citet{Mehta2011} attributed the excavation of Mars regolith by the Phoenix spacecraft to DGEE, where the impingement of supersonic jets fluidized the porous soil, leading to a large erosion rate.

In contrast, for lunar landings, \citet{Dotson2025a} and others have argued that the primary mechanism by which thrusters erode the surface is VE and the other mechanisms identified above are less influential due to the expected tightly packed nature and high-bearing strength of lunar regolith. 

Across different mechanisms and gravity regimes, the scaling behavior for gas excavated regolith is poorly understood, particularly as to how laboratory experiments and the results of landings on the Moon and Mars can be used to to better understand plume-surface interaction on asteroids. Unlike traditional impact experiments, crater formation through gas firing is far less studied, and it is not immediately clear whether thrusters can be used to reliably obtain an estimate of regolith geotechnical properties. In this study, we will clarify the extent to which spacecraft thrusters can be used as a robust tool for asteroid characterization. 

Here, we explore cratering from gas-based excavation with a focus on developing crater-scaling relationships for thruster-based excavation of asteroid regolith. The main driving mechanisms here are likley BCF, DDF, and DGEE. We present a series of experiments that were undertaken with the goal of constructing reliable models and scaling relationships for gas-based excavation in low-gravity environments. These experiments enable an essential understanding of how thrusters can be used to explore low gravity planetary surfaces. Models exist for how regolith reacts to thruster fire on larger bodies \citep{Metzger2009}, but little is known about how to extend such relationships to low-gravity airless bodies. The outcome of this study could be used to understand thruster-based excavation at (99942) Apophis \citep{DellaGiustina+2023}, and future small body rendezvous targets. In general, if the gas-excavation process can be better understood, rendezvous asteroid missions have the potential to use their on-board thrusters as a tool for understanding near-surface and interior properties of mission targets. 

%% file: _02_methods.tex
\section{Methods}
\label{sec:methods}
\subsection{Gas-Based Cratering Experiments:}

\begin{figure}
    \centering
    \includegraphics[width=\columnwidth]{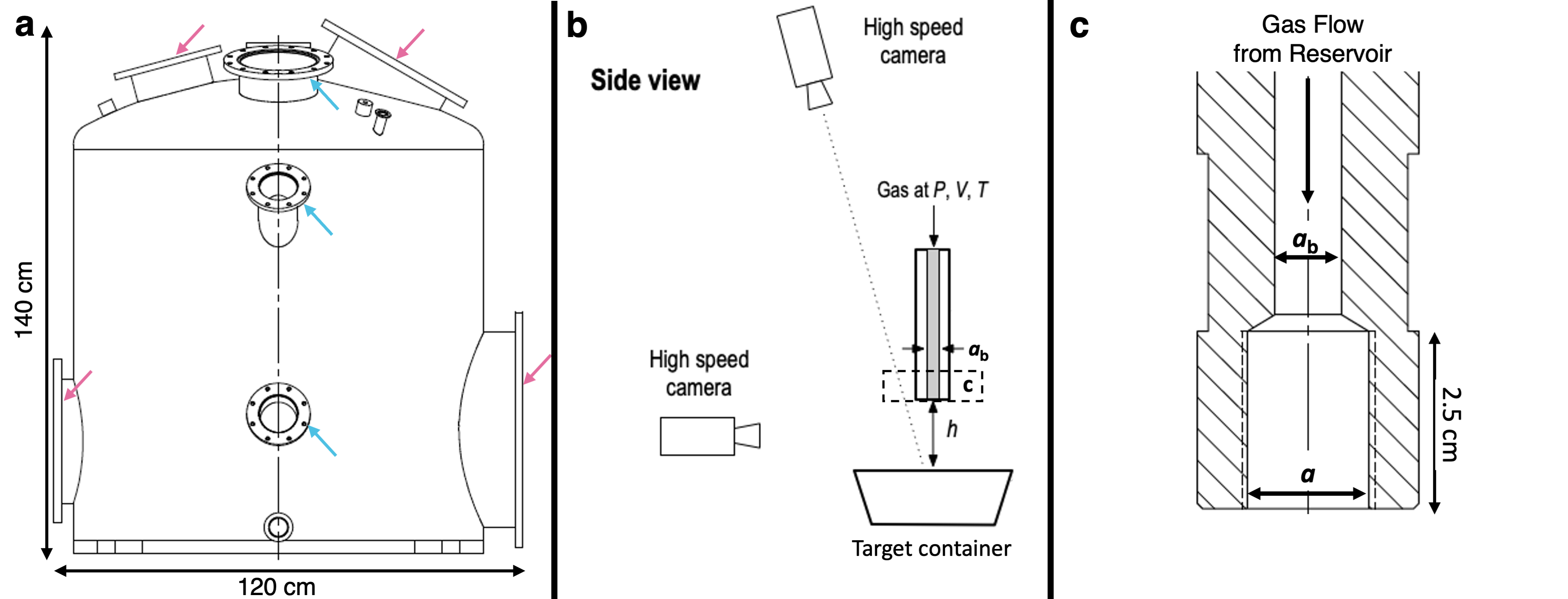}
    \caption{Gas excavation experiment apparatus and setup. \textbf{a,} Schematic of vacuum chamber within which gas excavation experiments are performed. Cyan arrows indicate ports at different angles relative to the center of the chamber floor where the barrel, connected to a gas reservoir, is inserted. Magenta arrows indicate viewing ports. \textbf{b,} a side view of the experimental setup. High speed cameras are place on the outside of the vacuum chamber viewing ports. Pressurized N$_2$ gas flows from a reservoir through an aluminum barrel and exits into the vacuum chamber, where we place a granular target at a vertical distance $h$ from the ``nozzle'' end of the barrel, highlighted by a dashed box and shown in more detail in \textbf{c}. \textbf{c,} a detailed schematic of the barrel ``nozzle'', which shows the inner diameter of the barrel, $a_b = 0.97$ cm, that expands out into a 2.5 cm-long cylindrical volume with a diameter, $a =$ 2.5 cm.}
    \label{fig:experimental_setup}
\end{figure}

\subsubsection{Experiments in $1g$ using APL Impact Lab's Light Gas Gun}  

We performed experiments at the Johns Hopkins University Applied Physics Laboratory’s Impact Lab (IL). For our gas-based cratering experiments, we reconfigured the IL gas gun which is typically used to perform low-speed impact experiments, which fires projectiles at speeds of 0.1 to 1 km/s, to perform gas-only “impact” experiments. The lab’s gun consists of a single-stage, compressed inert gas gun and a large vacuum chamber (Fig.~\ref{fig:experimental_setup}a). The vacuum chamber is 1.2 m in diameter and 1.4 m tall, and it is pumped down to an ambient pressure of $\sim$1 Torr before each experiment. The pressure inside the vacuum chamber is measured with a vacuum gauge that has a precision of 0.1 Torr. In its normal configuration, the light gas gun is setup by attaching a projectile to a sabot, and loading the sabot into the projectile-chamber, which is then sealed. The system is then pressurized with $N_2$ gas by opening a valve (valve 1), which allows gas to flow through approximately 1-m of plumbing until it reaches another valve (valve 2) close to the gas-loading chamber, which is behind the projectile chamber.  Valve 2 is opened to fill the gas-loading chamber with 32.8 cm$^3$ of $N_2$ gas at ambient temperature. The static pressure of the gas in the system is then modulated using a regulator to our chosen pressure, after which valve 2 is then closed. The gun is then fired, releasing the gas from the gas-loading chamber and into the projectile chamber, which drives the sabot into a 1-m-long barrel with an inner diameter of 0.95 cm. For projectile experiments, a sabot stopper is installed onto the down-stream end of the barrel, which "catches" the sabot and allows the projectile to continue along its trajectory into the vacuum chamber. For our experiments, we do not load a sabot into the projectile chamber, or use the sabot-stopper. Rather, the pressurized $N_2$ gas flows directly through the barrel and empties out into the vacuum chamber. Fig.~\ref{fig:experimental_setup}b shows our experimental setup.

\begin{figure}
    \centering
    \includegraphics[width=0.5\linewidth]{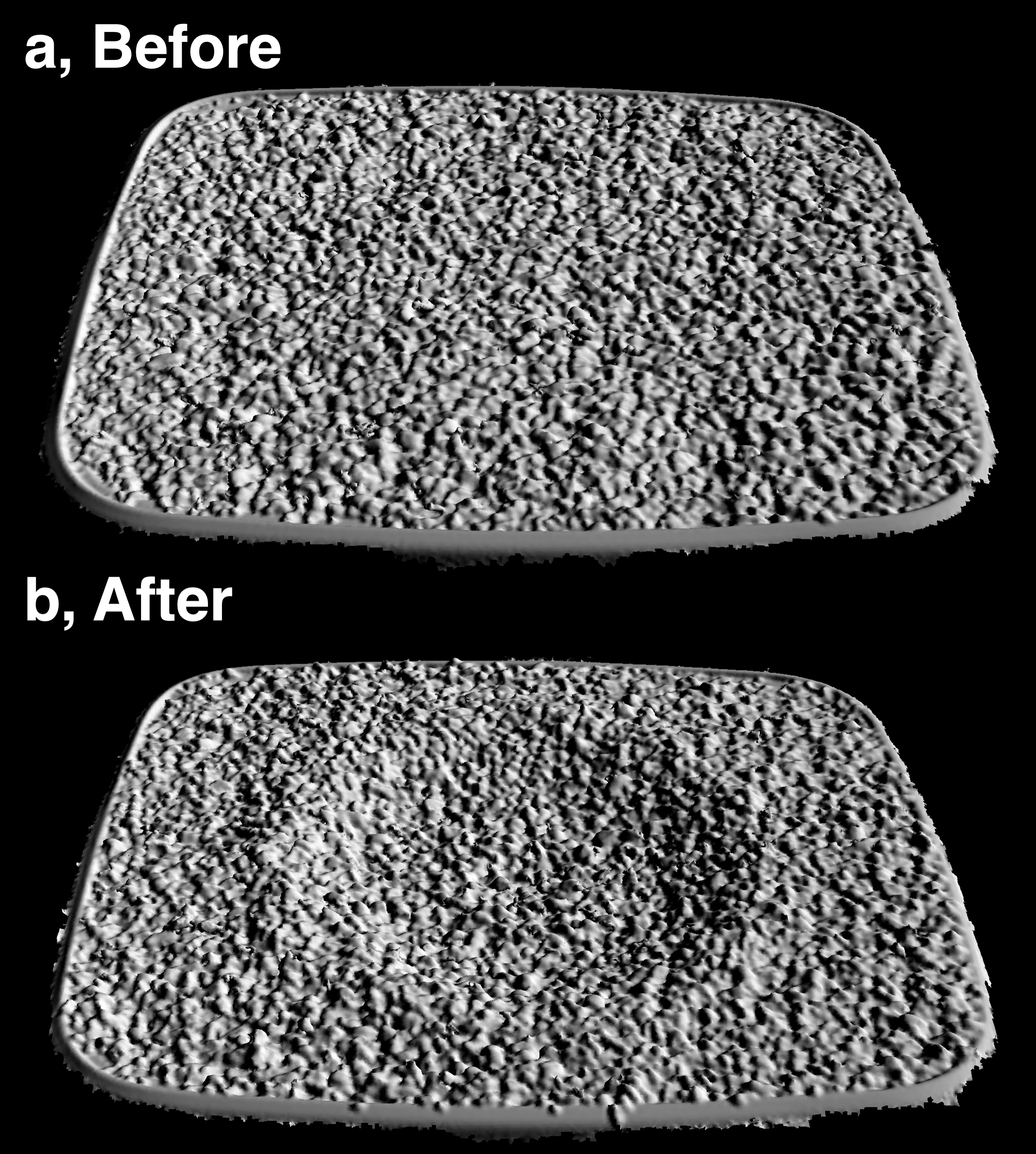}
    \caption{Crater volumes are measured with 3D scanners.}
    \label{fig:3D_scan}
\end{figure}

The gas properties (temperature, pressure, and volume) are measured or known upstream in the gas-loading chamber. As the gas is released, it expands through the barrel before venting into the vacuum chamber.  Since the reservoir volume is smaller than the volume of the barrel (and the vacuum chamber itself), it experiences unsteady compressible flow inside the pipe, leading to super-sonic flow before exiting the barrel. Additionally, the inner diameter of the barrel widens in the final 2.5-cm of the barrel as shown in Fig.~\ref{fig:experimental_setup}c, which leads to the acceleration of the gas to supersonic flow as it vents into the chamber. We simulated this effect using the computational fluid dynamics software \citet{ansys_fluent_2023} to understand the effect of this geometry on the exit properties of the gas and calculate the downstream values (see Appendix for model details).
As we detail later, estimating down-stream properties in this manner allows us to construct scaling relationships that are framed in gas conditions at a ``nozzles"-exit as these are typically well-defined for spacecraft thrusters.   
For our experiments, we lowered the barrel of the gun into the impact chamber until it was near the surface of the regolith target, varying the height between the end of the ``nozzle", $h = -1$--$10$ cm . Negative values of $h$ are achieved by lowering the nozzle such that it penetrates into the granular target. This was done to simulate the conditions in which the OSIRIS-REx TAGSAM penetrated into the surface of Bennu before sampling \citep{Lauretta2022, Walsh_2022}. We also systematically varied upstream gas pressure, $P_0$, which in turn modulated the exit gas density, $\delta$. The value of $\delta$ and the gas speed at the nozzle, $U$, was calculated using \citet{ansys_fluent_2023} (see Appendix). Table~\ref{tab:gas_experiment_conditions} summarizes the experiments that we conducted. 

The regolith targets were made of a cohesionless gravel with a bimodal particle distribution that is intended to mimic the properties of the surface material seen on Bennu \citep{Barnouin+2022}. The gravel has a peak diameter of 1.3 mm for 18\% of its incremental volume, and 4.4 mm for 32\% of its incremental volume. The full range of particle sizes spans from 0.07 to 8 mm. The gravel has a mean particle aspect ratio of 0.68, with the smaller particles being slightly more rounded (sphericity near 0.83) relative to the larger ones (sphericity near 0.8). The angular nature of the gravel is reflected in the measured angle of repose of the gravel, which is steep at 38$\pm$3$^\circ$. The bulk density of this material is 1,652$\pm$2 kg/m3; individual grains measure 2,394$\pm$2 kg/m3. Therefore, the target had a bulk porosity of approximately 31\%.

We observed the excavation of the resulting gas-generated craters using high speed videos (1200 to 1600 f/s) and a laser profile. These cameras allow us to qualitatively evaluate the growth of the gas crater.  Using an Artec3D Artec Eva handheld scanner, we scanned the regolith targets before and after each experiment. The scan data are used to generated before and after digital terrain models of the target at sub-mm accuracy in order to accurately measure the volume of displaced material  (Figure~\ref{fig:3D_scan}).

\input{Tables/Setup_Table}

\subsection{$\pi$-scaling Relationships}
Crater scaling relationships are an essential tool used for relating the properties of a i) projectile and ii) target to the properties of the iii) resulting crater, in particular the volume of a crater and the speed of ejecta. In general, crater scaling relationships can be used to estimate the properties of one of these three objects (the projectile, the target, or the crater) if the properties of the other two objects are known. 

Typically,crater scaling relationships are used to relate observations of crater dimensions on a planetary surface (the target) to the properties of the projectiles that formed those craters. For a population of projectiles, such as asteroids or comets, with well described statistical properties, such as relative speeds and densities, crater scaling relationships can be used to relate a crater's size to the size of the body that formed that crater. Thus, these scaling relationships become an essential tool for crater chronology studies, where a surface's exposure age can be determined based on its cratering record \cite[e.g.,][]{Melosh1989, Bottke+2005, Lunar_Cratering, Martian_Cratering, Mercury_Cratering, Bottke20202019je00}.  
In a similar manner, if an impactor's properties are fully understood, such as in experiments, then characterization of an impact crater's properties can provide a means to directly estimate a target's geotechnical properties, including its strength, bulk density and cohesion. Such an approach has been used to characterize the surface and sub-surface properties of: i) the Moon by the LCROSS mission \citep{LCROSS2010}, ii) the comet Tempel 1 by the Deep Impact mission \citep{AhearnDeepImpact}, iii) the asteroid Ryugu by the Hayabusa 2 mission \citep{arakawa_artificial_2020}, and iv) the asteroid by the DART mission \citep{Daly+2023}. Critical to this type of effort is the construction of a crater scaling relationship based on laboratory or numerical experiments using analog surface materials. 

A common formalism used  in the construction of a crater scaling relationship is one introduced by  \citet{Holsapple+1987}  based on the Buckingham-$\pi$ theorem. This formalism uses dimensionless numbers, called $\pi$-group parameters, to relate a crater's volume, \textit{V\textsubscript{C}}, to impactor and target properties such as the impactor radius, $r_p$ impactor velocity, $U_p$, impactor mass density, $\delta_p$, target density, $\rho$, target strength, \textit{Y}, and surface gravity, \textit{g.} In the formalism of \citet{Holsapple+1987}, the $\pi$-group parameters are related by

\begin{equation}
    \pi_V=K_1 [\pi_4^{-1/3}\pi_2+K_2(\pi_4^{-1/3}\pi_3 )^{(2+\mu)/2} ]^{-3\mu /(2+\mu )}, 
    \label{eq:pi_equation}
\end{equation}
and

\begin{equation}
\pi_V=\frac{\rho V_C}{m_p} = \frac{\rho(R_C /K_r )^3}{m_p}    ,
\label{eq:pi_v}
\end{equation}
where $m_p$ is the projectile mass, and $R_c$ is the crater radius. $K_1$, $K_2$, $K_r$, and $\mu$ are crater scaling constants that depend on the target material. $K_r$ relates the radius of the crater to its volume, which can be estimated through crater morphometric measurements.  The $\pi$-group parameters that control the cratering efficiency, $\pi_V$, are

\begin{equation}
\begin{aligned}
  \pi_2&=gr_p/U_p^2 , \\
  \pi_3 &=Y/(\rho U_p^2 ) , \\
  \pi_4&=\rho /\delta_p.
  \end{aligned}
  \label{eq:pi_terms}
\end{equation}

Crater scaling constants are determined through laboratory or numerical experiments for a given type of target material, by systematically varying the $\pi$-group parameters. This effectively allows for the ``construction'' of a crater scaling relationship that can be applied to the use-cases described above. For the case of a surface that has low strength compared to local gravity, the first term in the square brackets in Eq.~\ref{eq:pi_equation} dominates. This occurs when, 
\begin{equation}
\pi_2 \pi_4^{-1/3}  = K_2 (\pi_4^{-1/3}\pi_3 )^{(2+\mu)/2}.\\
\label{eq:pi_term_dominant}
\end{equation}
Substituting the $\pi$-terms from  Eq.~\ref{eq:pi_terms} into Eq.~\ref{eq:pi_term_dominant}, we obtain the following relationship for the value of $Y$ where cratering transitions from gravity-dominated outcomes to strength-dominated outcomes:
\begin{equation}
Y_\mathrm{tr} = K_2^{-\frac{2}{\mu + 2}} (g r_p)^{\frac{2}{\mu + 2}} U_p^{\frac{2\mu}{\mu + 2}} \rho^{\frac{4\mu + 6}{3(\mu + 2)}} \delta_p^{-\frac{\mu}{3(\mu + 2)}}
\label{eq:Y_tr}
\end{equation}

Prior to the recent exploration of small near-Earth asteroids Bennu and Ryugu by the OSIRIS-REx and Hayabusa2 missions, respectively, it was thought that the strength term would dominate as these bodies have microgravity environments \citep{Arakawa+2017}. Surprisingly, both missions showed that the surfaces of the asteroids are incredibly weak and dominated by their gravity. Here, we are primarily interested in using $\pi$-group parameters to relate thruster firing conditions and target properties to cratering outcomes for gravity-controlled surfaces. Therefore, our experiments on to granular targets allow us to probe this regime. For gas-based craters, we update the formalism by using gas properties in place of projectile properties, and introduce two dimensionless $\pi$-group that describe key properties of a PSI event. The first is a reworking of the $\pi_2$ parameter that takes into consideration gas density and target particle size, \textit{a}, that is based off of the densimetric Froude number, $F_r$ , introduced by \citep{RajaratnamBeltaos1977},
\begin{equation}
  \pi_{2,g}= F_r^{-2} = (\rho-\delta)gD/(\delta U^2) , \\
  \label{eq:pi_2_g}
\end{equation}
where $D$ is the mean diameter of a regolith particle, $\delta$ is the gas density, and $U$ is the gas velocity. \citet{RajaratnamBeltaos1977}  showed that the depth of excavation by a gas jet impinging on regolith reaches an asymptotic depth, $d_a$, that scales with $F_r$ and the ratio of the impingement height, $h$ , to the diameter of the jet at the nozzle, $a$. We also introduce a second $\pi$-group parameter that accounts for the thruster altitude so that the scaling relationship can capture the effect of gas expansion from the nozzle to the surface. 
\begin{equation}
  \pi_{h}=\left(\frac{h-a}{a}\right)^2 , \\
  \label{eq:pi_h}
\end{equation}
Unlike other formalisms of gas jet cratering that simply use the ratio of $h$ to $a$, we choose to take the difference in the numerator (Eq.~\ref{eq:pi_h}) to adequately capture cases where the impingement height is 0 (i.e., cases where the nozzle rests on the surface). Using this approach, we can capture cases similar to that of OSIRIS-REx's TAG maneuver at asteroid Bennu \citep{lauretta_spacecraft_2022,Walsh_2022,Bierhaus2026}, where the sampler's pressurized nitrogen gas was released below the pre-impact surface. The squared dependence on the height-to-nozzle-diameter ratio reflects the geometric spreading of the gas plume and the resulting variation in the surface area subjected to impingement. This scaling is motivated by the analytical expression for gas density as a function of distance from the nozzle exit \citep{Roberts1966}: 
\begin{equation}
    \frac{\delta}{\delta_e} = \frac{k}{8}\left(\frac{h}{a}\right)^{-2}\left(\cos\theta\right)^k,
\label{eq:roberts_density}
\end{equation}
where $\delta_e$ is the gas density at the nozzle exit, $\theta$ is the azimuthal angle from the nozzle center line, and $k$ is a constant that depends on the specific heat ratio and Mach number of the gas. We use the formulation of Eq. \ref{eq:pi_h} to avoid the non-physical limitation of Eq.~\ref{eq:roberts_density}, for small values of $h$,  when $h \leq a$ then $\delta>\delta_e$. But we introduce another issue for when $h \rightarrow a$ then $\pi_h \rightarrow 0$. We, remedy this  by adapting a minimum value of 1 for the ratio of height-to-nozzle-diameter, such that: 
\begin{equation}
  \pi_{h}=\max\left(1.0, \left(\frac{h-a}{a}\right)^2 \right). \\
  \label{eq:pi_h_max}
\end{equation}

For the case of gravity-dominated cratering by gas jets, as would be the case for cohesion-less granular materials, we propose the following scaling:
\begin{equation}
    \pi_{V,g}= \frac{\rho V_c}{m_g} = K_1 [\pi_{2,g} \pi_h \pi_{4,g}^{-1/3} ]^{-3\mu /(2+\mu )}, 
    \label{eq:pi_gas_equation}
\end{equation}
where $m_g$ is the total mass of the gas that impinges on the surface, and $\pi_{4,g} = \rho/\delta$.
Additionally, for the case of gas-jet cratering on cohesive soils, we introduce an additional strength term after \citet{Mehta2011}:
\begin{equation}
    \pi_{Y,g}= \frac{Y}{\delta U^2}, 
    \label{eq:pi_Y_gas_equation}
\end{equation}
such that a general crater-scaling relationship for gas jets could then be written as:
\begin{equation}
    \pi_{V,g}= \frac{\rho V_c}{m_g} = K_1 [\pi_{2,g} \pi_h \pi_4^{-1/3} +K_2(\pi_4^{-1/3}\pi_{Y,g} )^{(2+\mu)/2}]^{-3\mu /(2+\mu )}, 
    \label{eq:pi_gas_equation_general}
\end{equation}

In Section 3, we demonstrate the validity of these terms by comparing our experimental results with those of \citet{MasonNordmeyer1969}, who conducted a series of experiments using 1-lbf thrusters into cohesive lunar simulant.

%% file: Tables/Setup_Table.tex
\begin{deluxetable}{ccccc}
\tablecaption{Experimental conditions for gas cratering experiments conducted at APL's Impact Lab. $P_0$ is the upstream gas pressure; $\delta$ is the gas density at the nozzle exit; $P_{\mathrm{amb}}$ is the ambient pressure in the vacuum chamber; and $h$ is the distance from the nozzle exit to the regolith surface, measured with a $1\sigma$ precision of 1 mm.
\label{tab:gas_experiment_conditions}}

\tablehead{
    \colhead{Experiment} &
    \colhead{$P_0$ (psi)} &
    \colhead{$\delta$ (kg m$^{-3}$)} &
    \colhead{$P_{\mathrm{amb}}$ (psi)} &
    \colhead{$h$ (cm)}
}

\startdata
1  & 200  & 0.47  & 0.03 & 5  \\
2  & 200  & 0.47  & 0.03 & 0  \\
3  & 400  & 0.94  & 0.03 & 5  \\
4  & 400  & 0.94  & 0.03 & 0  \\
5  & 600  & 1.41  & 0.03 & 10 \\
6  & 600  & 1.41  & 0.03 & 5  \\
7  & 600  & 1.41  & 0.03 & 0  \\
8  & 800  & 1.88  & 0.02 & 10 \\
9  & 800  & 1.88  & 0.03 & 5  \\
10 & 800  & 1.88  & 0.03 & -1 \\
11 & 800  & 1.88  & 0.03 & 0  \\
12 & 1000 & 2.35  & 0.03 & 10 \\
13 & 1000 & 2.35  & 0.02 & 5  \\
14 & 1000 & 2.35  & 0.03 & -1 \\
15 & 1200 & 2.82 & 0.03 & 10 \\
16 & 1400 & 3.29 & 0.03 & 10 \\
\enddata
\end{deluxetable}

%% file: _03_results.tex
\section{Results}
\label{sec:results}

\begin{figure}
    \centering
    \includegraphics[width=0.8\linewidth]{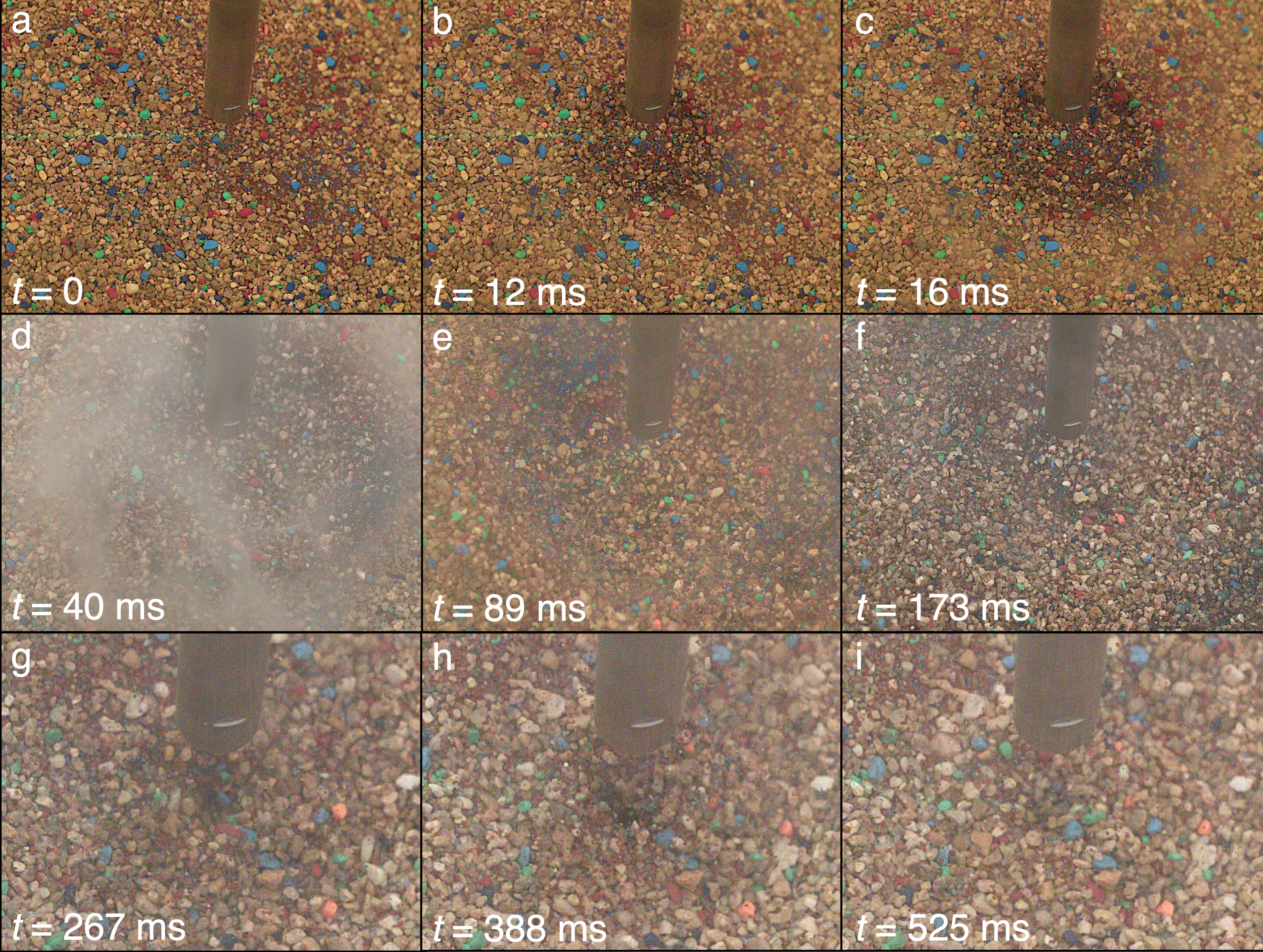}
    \caption{Experiment 13 captured at 1,200 fps from a window  slanted at 15-deg  from the long-axis of the gas barrel. The full movie is provided in the Supplementary file. The contrast and brightness of image frames have been adjusted to improve clarity over the experiment sequence. Full camera frame views of the experiments are shown in panels \textbf{a-f} and zoomed in views of the collapse of the stagnation region below the nozzle are shown in panels \textbf{g-i}. \textbf{a}, initial conditions of bed right before gas impinges on the surfaces. The time after gas impingement, $t$, is shown on the bottom left of each panel. \textbf{a}, initial conditions at $t = 0$. \textbf{b}, $t = 12$ ms, material begins to move radially outward from sub-nozzle point. \textbf{c,} $t = 16$ ms, a circular ring of dust can be seen surrounding the excavation area, the darker circle of particles. \textbf{d}, $t = 40$ ms, the dust ejection angle is now vertical and begins to obscure this view. \textbf{e}, the ejected dust begins to dissipate. \textbf{f}, $t = 173$ ms, the stagnation region, approximately the same size as the barrel diameter, begins to collapse. \textbf{g}, $t = 267$ ms, a zoomed in view of the stagnation region showing the formation of a pit. \textbf{h}, $t = 388$ ms, finer particles continue to eject from the central region. \textbf{i}, $t = 525$ ms, the central pit collapses and crater formation ends.}
    \label{fig:Top_Snapshots}
\end{figure}

\begin{figure}
    \centering
    \includegraphics[width=\linewidth]{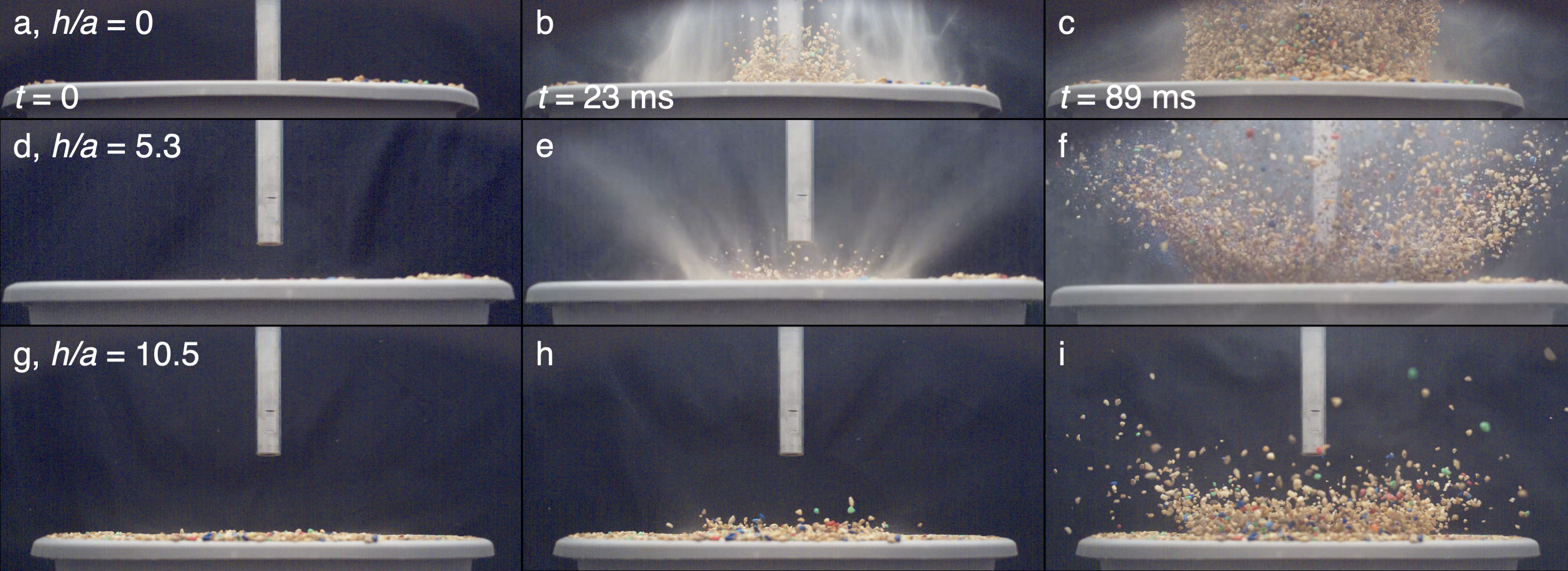}
    \caption{Experiment 12, 13, and 14 captured at 1,600 fps from a side view port-hole in the vacuum chamber. The full movie sequences is presented in Supplementary files. \textbf{a--c} Sequence Experiment 14, $h/a = 0$, at $t = 0$ (\textbf{a}), $t = 23$ ms (\textbf{b}), and $t = 89$ ms (\textbf{c}). \textbf{d--f} same as \textbf{a--c} but for Experiment 13, $h/a = 5.3$. \textbf{g--i} same as \textbf{a--c} but for Experiment 12, $h/a = 10.5$. }
    \label{fig:Side_Snapshots}
\end{figure}

In this section, we report on the outcomes of the 1$g$ gas cratering experiments, construct $\pi$-scaling relationships to describe these experiments, and evaluate the ability of the scaling relationship to predict the outcome of plume-regolith impingement experiments in the literature. As an illustrative example of an experimental outcome, Figs.~\ref{fig:Top_Snapshots} show sequence of snapshots from experiment 13, where the target and nozzle were separated by 5-cm and the gas was pressurized to 1,000 psi. The side views in Fig.~\ref{fig:Side_Snapshots}, shows experiments 12, 13, and 14 captured at a frame rate of 1600 frames per second (fps), highlighting the initial ejection of a dust cloud when the nozzle is close to the surface that can eject at a nearly vertical angle relative to the plane of the target.  Fig.~\ref{fig:Top_Snapshots} shows a top-down view of the same experiment, captured at a frame rate of 1200 fps. In all experiments, top-down views showed that the released gas peels layers of regolith away. This view shows how material is excavated, by moving radially outwards from the sub-nozzle point. A stagnation region develops directly below the nozzle exit that is approximately the same size as the nozzle diameter forming a pit. Finer particles continue to eject from this central region, until the central pit collapses during a crater modification stage. This behavior was typical of all experiments.  

\subsection{Gas Crater Morphological Properties}
\begin{figure}
    \centering
    \includegraphics[width=0.5\linewidth]{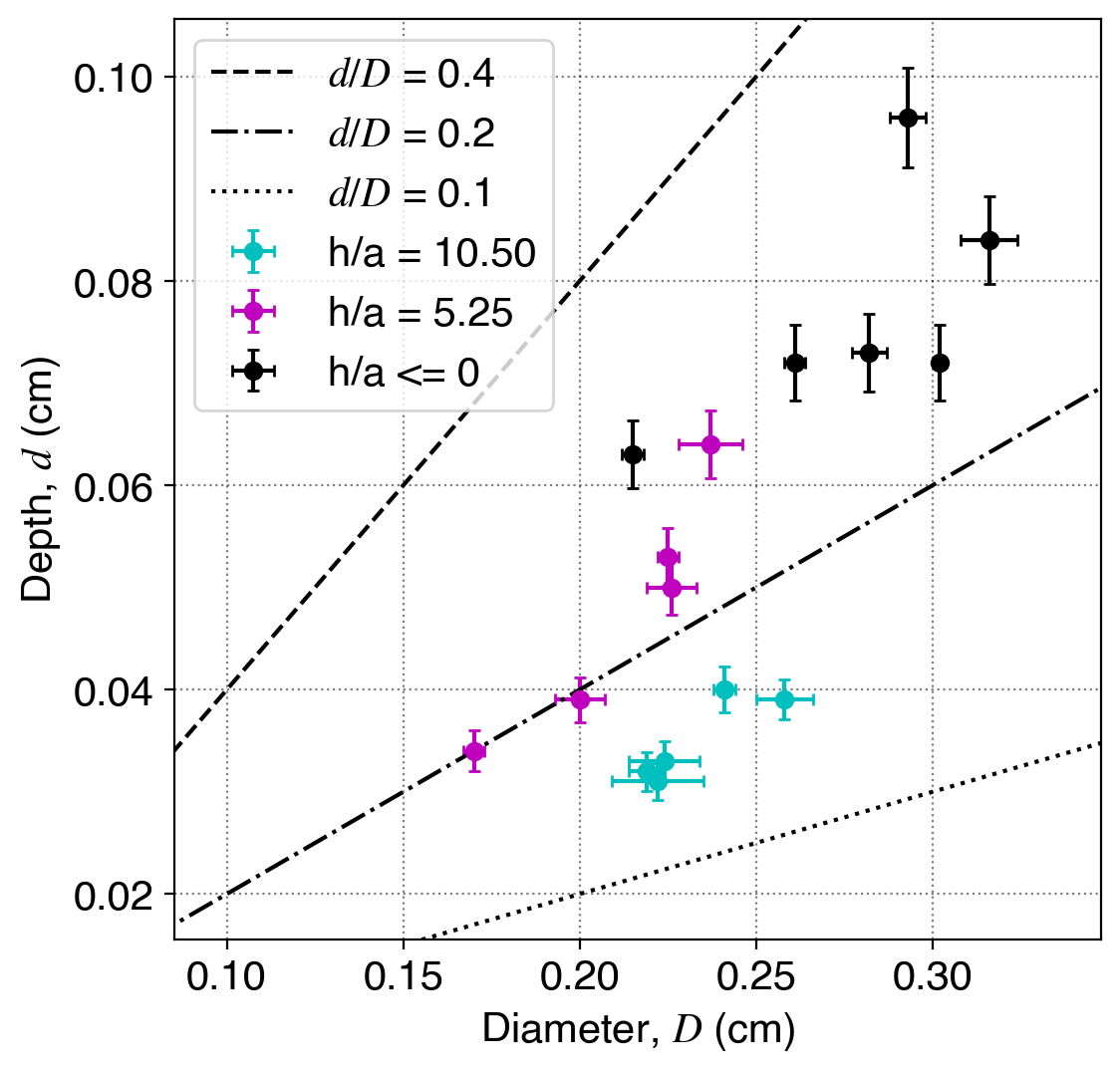}
    \caption{The depth, $d$, of gas craters compared to their diameters, $D$. The cyan, magenta, and black circles show measurements for craters where the normalized nozzle height, $h/a$, was 10.5, 5.25, and $\le$ 0, respectively. The dashed, dot-dashed, and dotted curves show the relationship between $d$ and $D$ for craters with $d/D =$ 0.4, 0.2, 0.1, respectively. The error bars are the 1-$\sigma$ measurement uncertainty.}
    \label{fig:crater_dD}
\end{figure}

In Table~\ref{tab:experiment_results}, we present the outcomes of the gas cratering experiments. For the range of conditions we explored, we found that craters had a  range of diameters, $D$ , from 17 cm to 32 cm, and depths, $d$ from 3.1 to 9.6 cm.  Fig.~\ref{fig:crater_dD} shows how crater depths vary with crater size with data points grouped by the normalized nozzle height, $h/a$, of the experiments. The $h = 0$ and $h = -1$ results are grouped together. It is evident from Fig.~\ref{fig:crater_dD} that the normalized nozzle height plays an important role in determining the depth-to-diameter ratio, $d/D$, of gas craters. We find that the craters with $h/a$ = 10.5, 5.25, and $\le$ 0 have mean $d/D = 0.15\pm0.01, 0.23\pm0.03,$ and $0.28\pm0.03$, respectively. The self-similarity in crater shapes is similar to that from projectile-based cratering, with the important caveat that height of impingement modulates the overall shape of the crater. 
A concern for cratering experiments into regolith, in general, is that the size of the container is insufficiently large to capture the full potential of crater growth. In the 1-g experiments, the container had a width of 60-cm and a depth of 19 cm, which is approximately twice as large as the largest and deepest craters. Combined with the observation of self-similar shapes for constant impingement height, we conclude that there was minimal influence of container edges on crater growth in these experiments.

\input{Tables/Results_Table}

\begin{figure}
    \centering
    \includegraphics[width=0.8\linewidth]{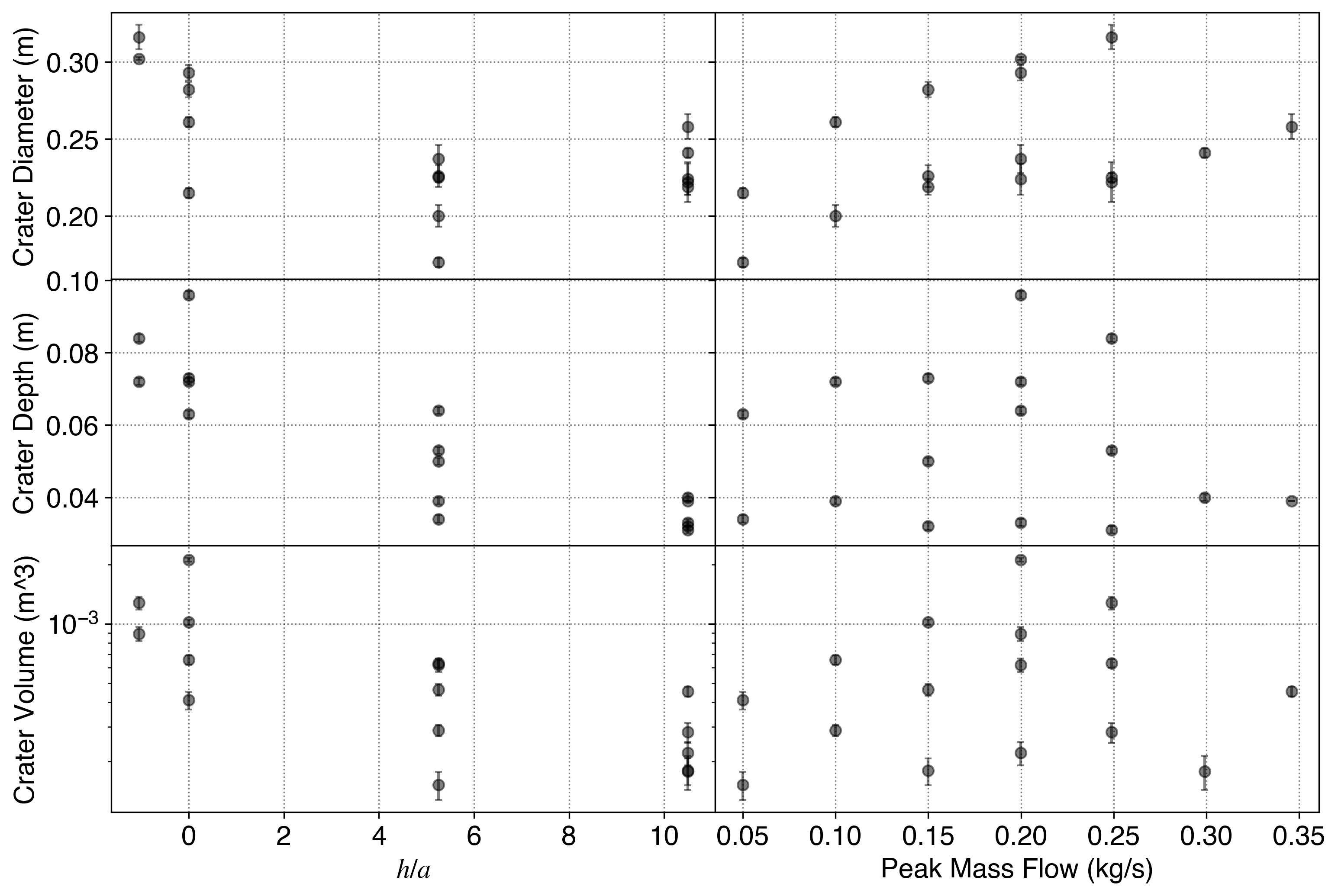}
    \caption{Crater morphometric properties as a function of the normalized impingement height, $h/a$, (left panels) and the peak mass flow of gas at the nozzle exit (right panels).}
    \label{fig:experiment_simple_results}
\end{figure}

In Fig.~\ref{fig:experiment_simple_results} we show how gas crater morphometric properties vary with the normalized impingement height, $h/a$ and the peak mass flow of gas at the nozzle exit. No clear relationship between crater and these gas properties emerges. The large scatter in results suggest that the regolith erosion is controlled by a combination of these factors and the properties of the regolith itself, which we explore in the next section.

\subsection{Gas Crater Volume Scaling Relationship in the Gravity Regime}

\begin{figure}
    \centering
    \includegraphics[width=0.6\linewidth]{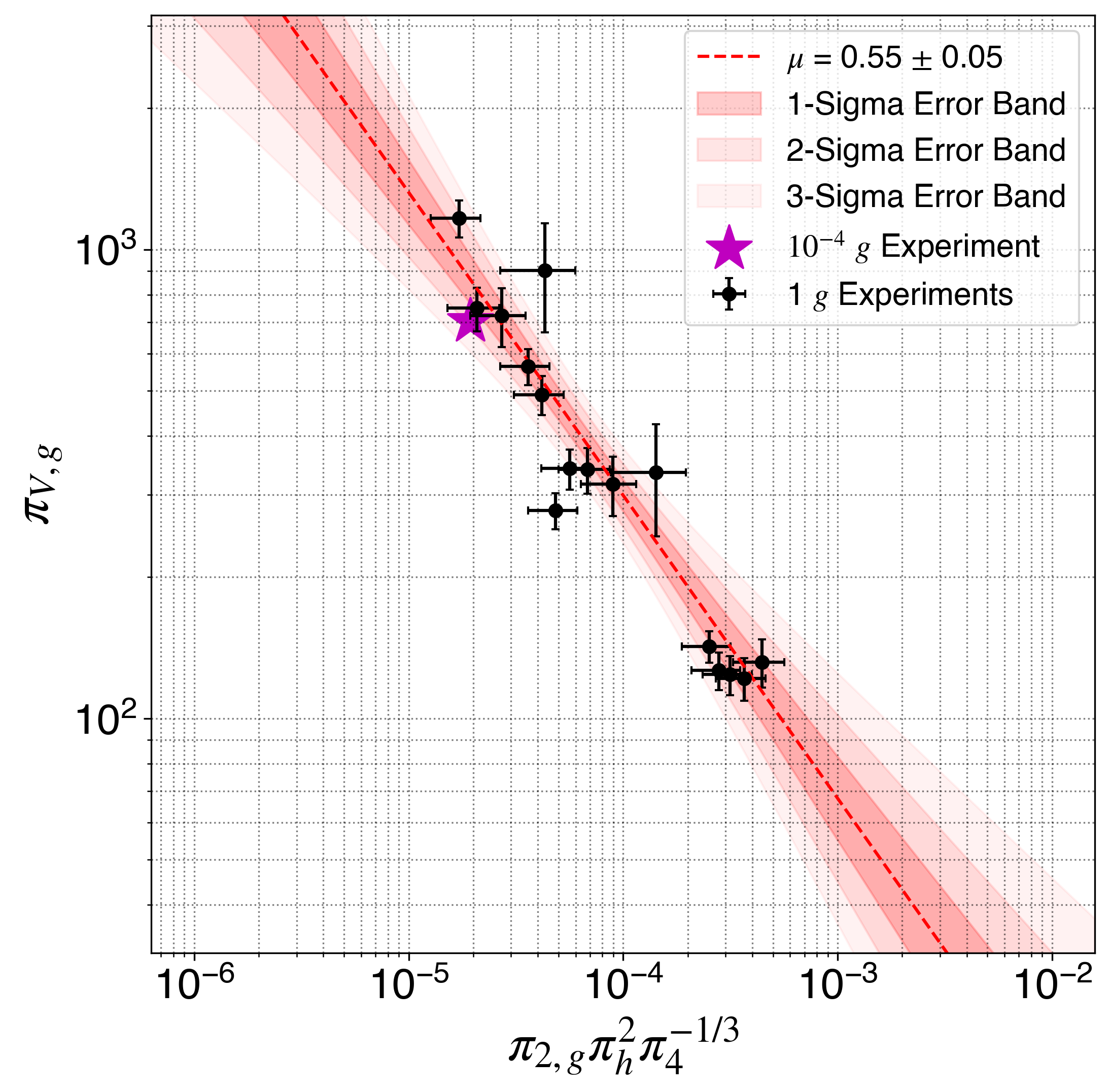}
    \caption{The $\pi$-numbers of gas cratering experiments are related by Eq.~\ref{eq:pi_equation}, shown as the red-dashed curve. The black dots show the results of the 1-g experiments, and the error bars are the propagated 1-$\sigma$ measurement uncertainty. The 1-, 2-, and 3-$\sigma$ error bands of the best-fit curve are shown in increasingly transparent red-shaded regions. The magenta star shows the result of the low-gravity drop-tower experiment by \citep{Baba2023}.}
    \label{fig:pi_scaling_gravity}
\end{figure}

We calculate the $\pi$-group numbers of each 1-g gas cratering experiment and perform a weighted least-squares fit to the relationship given by Eq.~\ref{eq:pi_gas_equation}, solving for the scaling constants $K_1$ and $\mu$. We choose this form of the equation, rather than the general equation because our experiments used cohesionless gravel. The result of this analysis is shown in Fig.~\ref{fig:pi_scaling_gravity}, and we calculate scaling constants $K_1 = 0.77\pm0.54$ and $\mu=0.55\pm0.05$, shown in the red dashed line. The large uncertainty in the value of $K_1$ is the result of the experiments having $\pi$-numbers with values such that  $X = (\pi_{2,g} \pi_h \pi_4^{-1/3})<<1$. Therefore, the uncertainty in scenarios with $X\sim1$ is large; however, the model can predict outcomes where $10^{-5} \lesssim X \lesssim 10^{-3}$ with more confidence. The value of $\mu$ is between momentum and energy scaling for gas-based excavation of regolith, as $V_c \propto m_g U^{1.3}$. Thus, the 1-g PIL experiments suggest that a gas crater volume scaling relationship in the gravity regime can be given by:

\begin{equation}
        \pi_{V,g} = 0.78 (\pi_{2,g} \pi_h \pi_4^{-1/3})^{-0.65}, 
\label{eq:pi_result_fit}
\end{equation}

\subsection{Low Gravity Gas Cratering Experiment}
\begin{figure}
    \centering
    \includegraphics[width=0.6\linewidth]{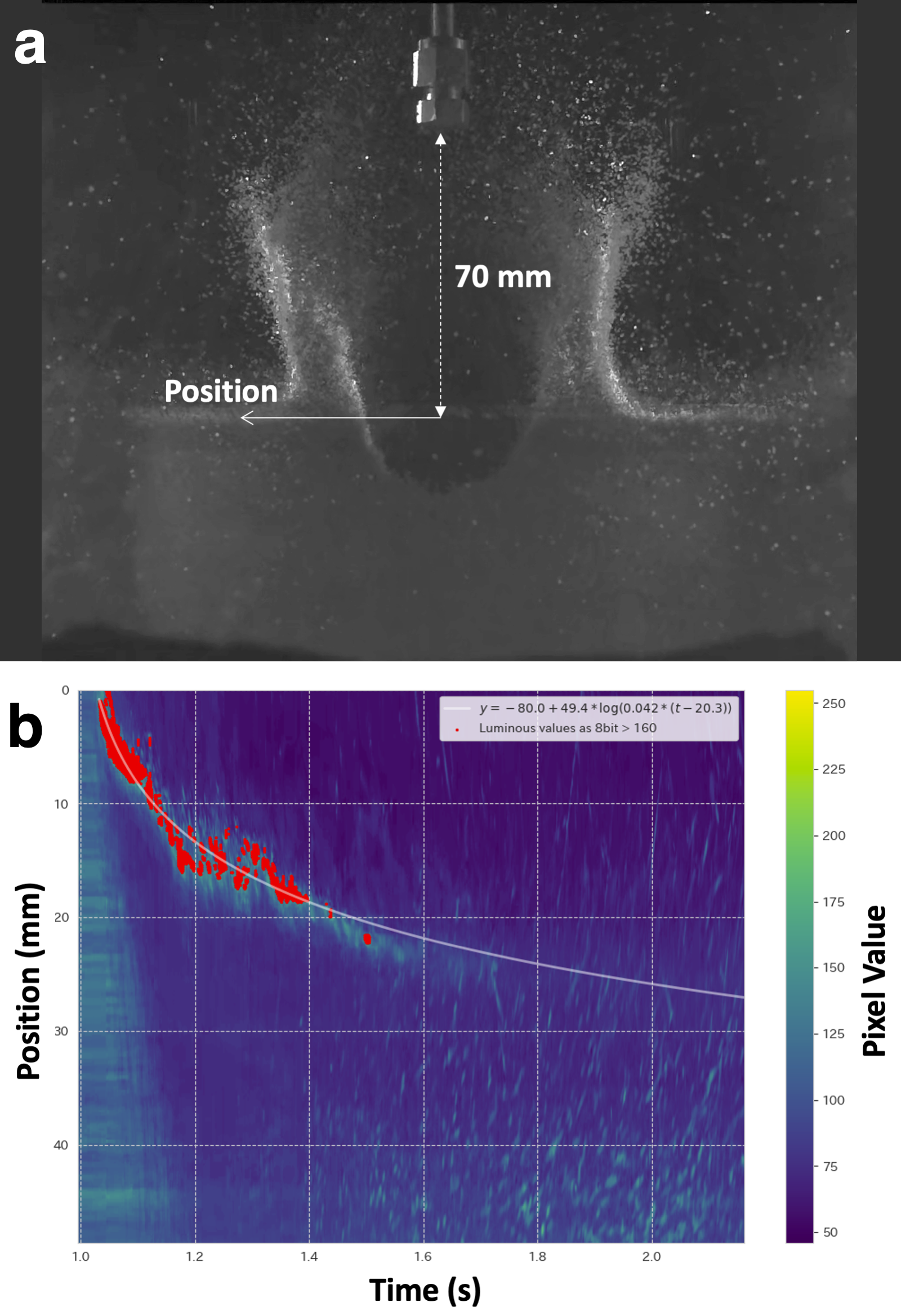}
    \caption{Low gravity ($\sim 10^{-4}g$) gas cratering experiment by \citep{Baba2023}. \textbf{a,} Snapshot of the experiment captured by a high speed camera. \textbf{b}, growth of the gas crater as a function of time is tracked by plotting the pixel value along the dimension shown by the horizontal white line in \textbf{a}. The red dots are detections of the crater edge (pixel values above a threshold), and the crater edge position ($y$) can be described by the logarithmic function of time ($t$) shown in the legend.}
    \label{fig:drop_tower_crater}
\end{figure}

To evaluate the generality of our proposed gravity-regime gas crater volume scaling relationship, we considered drop-tower experiments that were reported \citet{Baba2023}. In that work, a gas cratering experiment was installed in a capsule, which was dropped from the top of the COSMOTORRE facility in Hokkaido, Japan. During free fall, the gas cratering experiment inside the capsule is triggered, and a gas crater is formed under a gravitational acceleration of $\sim3\times10^{-3}$ m/s$^2$ (similar to the surface of $\sim$10-km asteroids). We used the JA-5V servo accelerometer to measure this low-gravity value. The experimental setup and procedures are described in detail in \citet{Baba2023}. Here, we briefly summarize key details. 
The drop-tower experimental setup consisted of three stages. The bottom stage contained a tank storing compressed air, and the center stage is equipped with a vacuum chamber. The vacuum chamber is equipped with a solenoid valve inside, which prevents air from flowing into the chamber before dropping. The third stage consists of a buffer tank connected to the vacuum chamber and an electric control unit that controls the operation. The second stage chamber is divided into two halves by acrylic plates with sharp knife-edge, one half of which is filled with the granular target, Tohoku silica sand\#5, which have particle density of 2590 kg/m$^3$, $\rho =$ 1320 kg/m$^3$ and $D = $ 0.53 mm. A nozzle extending from the manifold is placed directly above the target. A sheet laser is positioned to illuminate the line of this knife-edge. During the experiment, gas excavates a crater and particles moving in the plane of the sheet laser are illuminated. The crater formation process was captured by a high-speed camera at 3000 fps (Fig.~\ref{fig:drop_tower_crater}a). The nozzle was placed at $h = 70$ mm and had $a = 0.5$ mm. Upstream temperature and pressure sensors allow us to calculate gas speed and the mass flow rate of gas at the nozzle exit. 

For every image frame, we analyzed the pixel value along 1 dimension that originates from the plume impingement point directly below the nozzle (horizontal arrow in Fig.~\ref{fig:drop_tower_crater}a). Fig~\ref{fig:drop_tower_crater}b shows the result of this analysis, where the the growth of the crater edge is captured and can be described by a logarithmic curve (shown in the legend). The value of the time axis in Fig~\ref{fig:drop_tower_crater}b is measured after the drop of the capsule. The gas is released 1.0 s after the drop automatically, based on a sequencer-based trigger to wait when the low-gravity is achieved. The volume of the crater is then estimated as a half-sphere with a radius equal to the gas crater's position at a time, $t =$ 1.6 s (equivalent to 0.6 s after gas release). $m_g$ is then calculated based on the average mass flow rate multiplied by 0.6. 

We calculated the $\pi$-group numbers for the low-gravity drop-tower experiment and show it in Fig.~\ref{fig:pi_scaling_gravity} as a magenta star. The result is within the 1-sigma error band of Eq.~\ref{eq:pi_result_fit}. Though this is only a single experiment, we choose to show it here as the drop-tower experiments are challenging to conduct, yet extremely valuable to understand how PSI may by influenced by gravity. The overlap in the data point with our proposed $\pi$-scaling relationship provides supporting evidence for its validity across gravity regimes, which is crucial for understanding the outcome of thruster PSI on other planetary surfaces.

\subsection{General Gas Crater Volume Scaling Relationship}

Using the estimated values of the scaling constants, a general gas crater volume scaling relationship that includes both strength and gravity terms can now be written as,
\begin{equation}
    \pi_{V,g}= 0.77 [\pi_{2,g} \pi_h \pi_4^{-1/3} + K_2(\pi_4^{-1/3}\pi_{Y,g} )^{1.275}]^{-0.65}, 
    \label{eq:pi_gas_general_fit}
\end{equation}
with the remaining unknown quantity being the strength-scaling constant, $K_2$. The projectile-cratering experimental literature shows that $K_2$ is generally of order unity \citep{Holsapple2012, Nakamura2017}, though it is found to vary by a factor of a few depending on the target. To evaluate the validity of this formulation for cohesive regolith, we performed an extensive review of the PSI literature, and found that the results of \citet{MasonNordmeyer1969} were unique, in that these authors 1) experimented with cohesive simulant, 2) used chemical thrusters with thrust levels similar to that of the OSIRIS-REx ACS thrusters, and 3) directly measured crater volumes after the end of the crater formation process. \citet{MasonNordmeyer1969} reported the outcomes of nine PSI experiments performed in a vacuum chamber at approximately $10^{-4}$~torr using two chemical rocket engines: a small engine with a 34.5 cm nozzle-exit diameter operated at 5-18 N, and a larger engine with a 138.7 cm nozzle-exit diameter operated at 300-450 N. The target was a poorly sorted mixture of 25\% screened cinder, 50\% sand, and 25\% pumice, with a $D = $ 0.26 mm, and a cohesive strength, $Y = $ 3.4 kPa. The experiments were performed in a large 41-ft diameter vacuum sphere. Each engine firing lasted 0.5 s. \citet{MasonNordmeyer1969} directly provide the thrust, nozzle height, nozzle-exit diameter, and post-test crater volume for each experiment. However, the authors did not report nozzle-exit properties for mass flow rate, gas velocity, and gas density. We estimated these quantities using a Surveyor Vernier engine as a historical analog. The final Surveyor Vernier design had an average specific impulse of approximately 280 s and a throttling range of 133-463 N \citep{JPLSurveyorFinal1969}.  Based on this impulse, we adopt an effective exhaust velocity, $U = I_{\mathrm{sp}}g_0  = 2.75$ km/s. We also apply this value to the smaller 5-18 N engines. This extrapolation is supported by contemporaneous NASA tests of bipropellant thrusters of a similar class \citep{Rollbuhler1968ReactionControl,Rollbuhler1968Fractional}. Furthermore, multiplying the ratio of the exit areas of the two nozzles (0.062) to an average thrust of $\sim$300 N for the large engine gives 19 N (close to the upper end range of thrusts for the smaller engine). For each experiment, the nozzle-exit area, $A_e=\frac{\pi a^2}{4}$. For a given thrust, $F$, the reconstructed mass flow rate and momentum-equivalent gas density, $\delta$, are then, 
\begin{equation}
 \dot{m} =\frac{F}{U},
 \qquad
 \delta=
 \frac{\dot{m}}{A_e U}
 =\frac{F}{A_e U^2}.
\label{eq:mason_effective_properties}
\end{equation}
where $\dot{m}$ is the mass flow rate of the thrusters. Table~\ref{tab:mason_nordmeyer_reconstructed} summarizes the reconstructed parameters and their associated published crater volumes and thruster configurations.

\begin{deluxetable*}{ccccccc}
\tablecaption{Outcomes and reconstructed parameters for thruster
PSI experiments reported by \citet{MasonNordmeyer1969}.
\label{tab:mason_nordmeyer_reconstructed}}

\tablehead{
    \colhead{Experiment} &
    \colhead{$F$ (N)} &
    \colhead{$h$ (mm)} &
    \colhead{$a$ (mm)} &
    \colhead{$V_c$ (cm$^3$)} &
    \colhead{$\dot{m}$ (kg s$^{-1}$)} &
    \colhead{$\delta$ (kg m$^{-3}$)}
}

\startdata
MN-1 & 5.3   & 68.58  & 34.5  & 89.47     & $1.94\times10^{-3}$ & $7.53\times10^{-4}$ \\
MN-2 & 10.2  & 68.58  & 34.5  & 226.80    & $3.72\times10^{-3}$ & $1.44\times10^{-3}$ \\
MN-3 & 14.7  & 76.20  & 34.5  & 302.67    & $5.34\times10^{-3}$ & $2.07\times10^{-3}$ \\
MN-4 & 16.0  & 68.58  & 34.5  & 412.30    & $5.82\times10^{-3}$ & $2.26\times10^{-3}$ \\
MN-5 & 16.0  & 137.16 & 34.5  & 189.76    & $5.82\times10^{-3}$ & $2.26\times10^{-3}$ \\
MN-6 & 17.8  & 276.86 & 34.5  & $\approx0$ & $6.47\times10^{-3}$ & $2.51\times10^{-3}$ \\
MN-7 & 311.4 & 139.70 & 138.7 & 4298.33   & $1.13\times10^{-1}$ & $2.73\times10^{-3}$ \\
MN-8 & 311.4 & 304.80 & 138.7 & 2934.92   & $1.13\times10^{-1}$ & $2.73\times10^{-3}$ \\
MN-9 & 444.8 & 139.70 & 138.7 & 5791.19   & $1.62\times10^{-1}$ & $3.89\times10^{-3}$ \\
\enddata

\tablecomments{
$F$ is the engine thrust;
$h$ is the nozzle height above the surface;
$a$ is the nozzle diameter;
$V_c$ is the crater volume;
$\dot{m}$ is the estimated gas mass flow rate;
and $\delta$ is the estimated gas density at the nozzle exit.
Gas properties were reconstructed using the assumptions described
in the text. All experiments had a reported firing duration
of $0.50\pm0.05$ s.
}

\end{deluxetable*}

\begin{figure}
    \centering
    \includegraphics[width=0.8\linewidth]{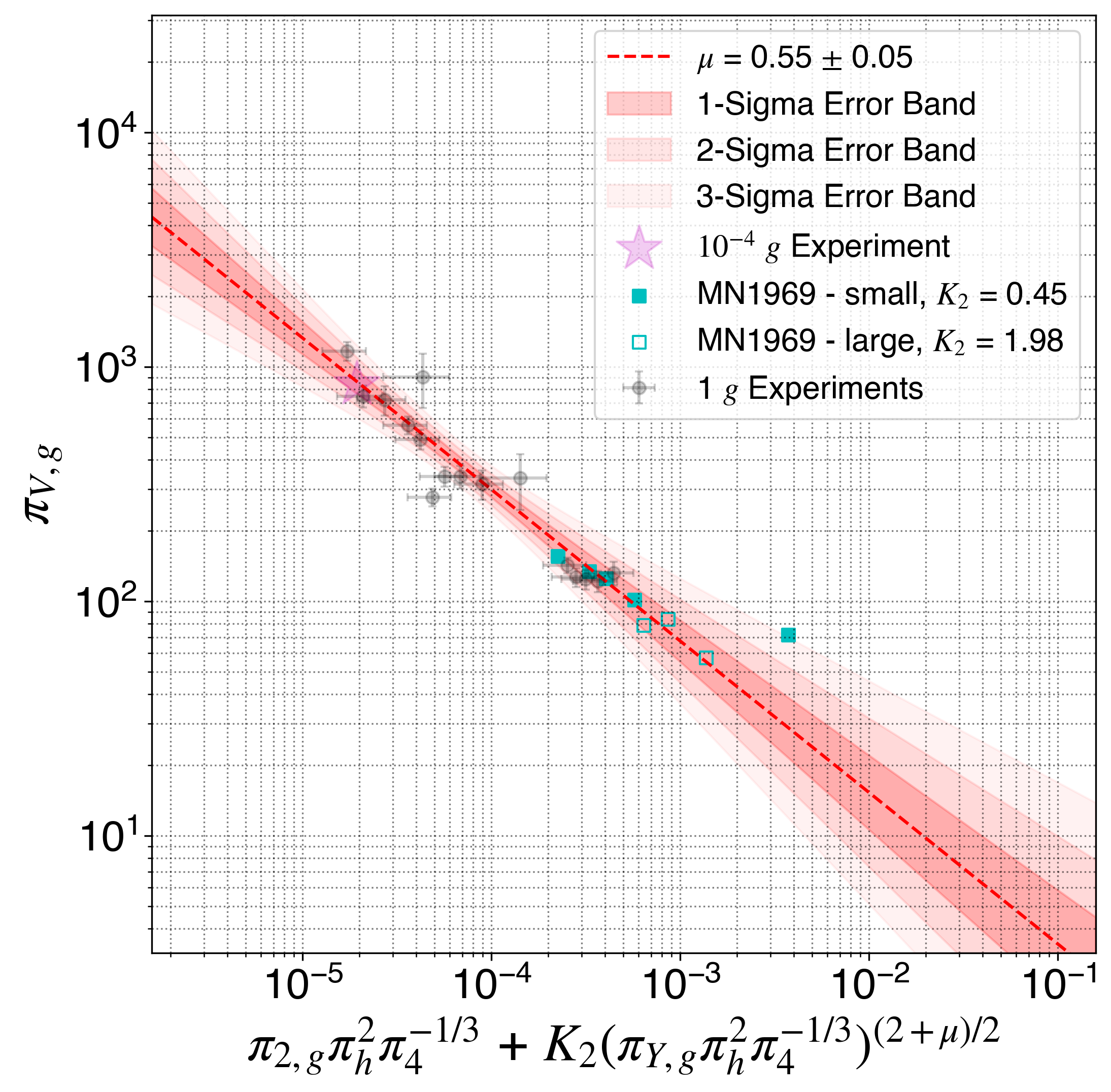}
    \caption{Outcomes of thruster PSI experiments by \citet{MasonNordmeyer1969} (cyan squares, MN1969) are well described by Eq.~\ref{eq:pi_gas_equation_general} (red dashed line) for values of $K_2 = 0.45$ and $1.98$ for small 5-18 N thrusters and large 300-450 N thrusters, respectively. Experimental values presented in this work are plotted for reference (black circles and magenta star). The 1-, 2-, and 3-$\sigma$ error bands of the Eq.~\ref{eq:pi_gas_equation_general} are shown in increasingly transparent red-shaded regions.}
    \label{fig:pi_scaling_general}
\end{figure}

We calculate the $\pi$-group numbers for these data and separate them by the relative size of the thrusters: small thruster cases for the 5-18 N experiments, and large thruster cases for the 300-450 N experiments. We did not consider the case of MN-6, which led to no measurement of the crater volume. We find the experimental outcomes of \citet{MasonNordmeyer1969} are best described by the formulation we have derived from our 1-$g$ experiments for values of $K_2 = 0.45$ and $1.98$ for cases with the smaller 5-18 N thruster and larger 300-450 N thruster, respectively (Fig.~\ref{fig:pi_scaling_general}). These values of $K_2$ are estimated by performing a cross-validation linear-regression of the \citet{MasonNordmeyer1969} data to Eq.~\ref{eq:pi_gas_equation_general},with $K_2$ as a free parameter. The coefficients of determination are 0.81 and 0.63 for the small and large thruster cases, respectively. 

This exercise demonstrates that our formulation $\pi$-scaling relationship can be applied to predict the outcome of PSI from the chemical thrusters. Though, it is more successful for the case of small thrusters with thrusts of 5-18 N. For the larger thrusters, the outcomes are within the $1-\sigma$ uncertainty of our model, but require large values for $K_2$ that are not typically observed for projectile-based $\pi$-scaling relationships.   

%% file: Tables/Results_Table.tex
\begin{deluxetable}{ccccccc}
\tablecaption{Results of gas cratering experiments.
\label{tab:experiment_results}}

\tablehead{
    \colhead{Experiment} &
    \colhead{$V_C$ (m$^3$)} &
    \colhead{$\sigma(V_C)$ (m$^3$)} &
    \colhead{$d$ (m)} &
    \colhead{$\sigma(d)$ (m)} &
    \colhead{$D$ (m)} &
    \colhead{$\sigma(D)$ (m)}
}

\startdata
1  & 1.52E-04 & 2.5E-05 & 0.034 & 0.001 & 0.170 & 0.003 \\
2  & 4.09E-04 & 4.2E-05 & 0.063 & 0.001 & 0.215 & 0.003 \\
3  & 2.87E-04 & 1.9E-05 & 0.039 & 0.001 & 0.200 & 0.007 \\
4  & 6.56E-04 & 3.5E-05 & 0.072 & 0.001 & 0.261 & 0.003 \\
5  & 1.79E-04 & 2.8E-05 & 0.032 & 0.001 & 0.219 & 0.005 \\
6  & 4.62E-04 & 3.2E-05 & 0.050 & 0.001 & 0.226 & 0.007 \\
7  & 1.02E-03 & 3.3E-05 & 0.073 & 0.001 & 0.282 & 0.005 \\
8  & 2.21E-04 & 3.0E-05 & 0.033 & 0.001 & 0.224 & 0.010 \\
9  & 6.19E-04 & 4.8E-05 & 0.064 & 0.001 & 0.237 & 0.009 \\
10 & 8.90E-04 & 7.4E-05 & 0.072 & 0.001 & 0.302 & 0.001 \\
11 & 2.12E-03 & 4.9E-05 & 0.096 & 0.001 & 0.293 & 0.005 \\
12 & 2.81E-04 & 3.3E-05 & 0.031 & 0.001 & 0.222 & 0.013 \\
13 & 6.30E-04 & 3.1E-05 & 0.053 & 0.001 & 0.225 & 0.003 \\
14 & 1.28E-03 & 9.6E-05 & 0.084 & 0.001 & 0.316 & 0.008 \\
15 & 1.78E-04 & 3.5E-05 & 0.040 & 0.001 & 0.241 & 0.003 \\
16 & 4.53E-04 & 2.6E-05 & 0.039 & 0.000 & 0.258 & 0.008 \\
\enddata

\tablecomments{
$V_C$ is the crater volume;
$d$ is the crater depth;
$D$ is the crater diameter.
The quantities $\sigma(V_C)$, $\sigma(d)$, and $\sigma(D)$
represent the corresponding $1\sigma$ uncertainties.
}

\end{deluxetable}

%% file: _04_discussion_conclusions.tex
\section{Conclusions and Future Work}

We investigated gas-driven crater excavation as a potential method for characterizing the regolith of low-gravity bodies using spacecraft thrusters. Table~\ref{tab:pi_scaling_summary} summarizes the different sets of experiments we analyzed here. We conducted pulsed N$_2$ experiments at $1$-$g$ in near-vacuum conditions using a cohesionless, Bennu-like gravel, while varying the upstream gas pressure and nozzle height. Across the conditions tested, the resulting craters ranged from 17 to 32~cm in diameter and from 3.1 to 9.6~cm in depth. The normalized nozzle height, $h/a$, systematically influenced crater morphology: the mean depth-to-diameter ratio increased from $0.15\pm0.01$ at $h/a=10.5$ to $0.28\pm0.03$ for $h/a\leq0$. By contrast, neither nozzle height nor peak mass flow alone described the full variation in crater size, demonstrating that gas-cratering outcomes depend on the coupled gas, target, gravitational, and geometric properties of the interaction.

\input{Tables/Discussion_Summary_Table}

We adapted the conventional projectile-cratering $\pi$-scaling framework to gas excavation by expressing crater volume in terms of the total delivered gas mass, gas density and speed, target bulk density and grain size, surface gravity, and nozzle height. A weighted least-squares fit to the cohesionless $1$-$g$ experiments yielded $K_1=0.77$ and $\mu=0.55$ (see Eq.~\ref{eq:pi_gas_equation}). The fitted value of $\mu$ is between momentum and energy-scaled excavation, giving $V_c\propto m_g U^{1.3}$. A low-gravity drop-tower experiment conducted at $g\sim3\times10^{-3}$~m~s$^{-2}$ falls within the 1-$\sigma$ uncertainty of the gravity-regime relationship. This agreement provides encouraging evidence that the proposed formulation can connect Earth-gravity experiments to asteroid-like gravity conditions, although the comparison is presently based on a single low-gravity experiment and should therefore be regarded as supporting evidence rather than a definitive validation. The $\pi$-scaling framework is further supported by the finite coupling region of the source (i.e., the input gas) and the target, with 1) the impingement region $<<$ gas crater radius, and 2) the timescale for gas venting from the nozzle ($\sim20$ ms) $<<$ the timescale for crater formation ($\sim500$ ms). While these conditions are met for our gas-cratering experiments, this may not be true for scenarios where 1) thrusting occurs for durations similar to the timescale for crater growth, and 2) where the gas dynamic pressure at the surface is lower than the surface's compressive strength. In these scenarios, viscous erosion may become the dominant mode of regolith particle mobilization. Future work will evaluate this transition between erosion modes.   

To extend the formulation to cohesive targets, we introduced an additional strength term and compared the relationship with the chemical-thruster experiments of \citet{MasonNordmeyer1969}. Holding $K_1$ and $\mu$ fixed at the values derived from the $1$-$g$ experiments, the small-thruster cases were best described by $K_2=0.45$ and had a coefficient of determination of 0.81. The large-thruster cases required $K_2=1.98$ and had a lower coefficient of determination of 0.63. Although the large-thruster outcomes remain within the 1-$\sigma$ uncertainty of the model, their comparatively high $K_2$ value and weaker fit indicate that extrapolation to larger thrusters is less well constrained. Additional experiments with independently measured nozzle-exit properties and target strengths are needed before $K_2$ can be confidently transferred across target materials and thruster classes.

Taken together, these results provide a preliminary crater-volume scaling framework. The framework offers a path toward predicting crater volumes from known maneuver and surface conditions and, ultimately, toward using observed thruster crater dimensions to constrain otherwise poorly known regolith properties. The present calibration remains limited in gravity, target strength, firing duration, and thruster class, but it supports the broader concept that spacecraft thrusters could serve as active geotechnical tools during future small-body reconnaissance missions.

Further work is needed to determine the extent to which thruster-based cratering can be used to characterize an asteroid's physical properties in an operational scenario. In particular, we identify the following knowledge gaps:

\textbf{1) Sensitivity to target properties.} Future experiments should use a controlled and replicated test matrix in which the characteristic grain diameter, $d_g$, target cohesive strength, $Y$, and porosity, $\Phi$, are varied systematically. Grain diameter should be varied over a sufficiently broad range to determine whether particle-scale entrainment, permeability, and the ratio of grain size to crater size introduce additional scale dependence. Target strength should be measured independently and varied to characterize the transition between gravity- and strength-controlled excavation. Porosity should be determined from the particle and bulk densities and varied to quantify its influence on gas penetration, compaction, and momentum transfer to the granular bed. \citet{Mehta2011} highlighted the role of regolith permeability in enhancing cratering efficiency in PSI through DGEE, by enabling the development of large pressure gradients within the target.  Eq.~\ref{eq:pi_gas_equation} does inherently contain a porosity dependence through the bulk density, $\rho = (1-\Phi)\rho_p$, where $\rho_p$ is particle density. As such, we find that the crater volume, $V_c \propto (1-\Phi)^{-1.43}$. Therefore, Eq.~\ref{eq:pi_gas_equation} predicts that plume impingement into regolith with $\Phi = 40-70\%$ leads to $2.1-5.6\times$ more volumetric erosion, compared to a surface with $\Phi=0\%$. These measurements would test whether such a scaling for $\Phi$ is to be expected, and, generally, whether a common set of $\mu$ and $K_1$ values can describe excavation across different target materials or whether additional dimensionless groups or material-dependent scaling constants are required.

\textbf{2) Long-duration thrusting.} The experiments presented here used a discrete volume of pressurized gas to form craters in granular material. In an operational scenario, a spacecraft thruster could fire for a longer duration set by the $\Delta v$ required for a near-surface maneuver. Future experiments should better quantify the effect of long-duration thrusting on gas cratering efficiency. Early work on crater formation from impinging gas and fluid jets suggested that an asymptotic crater that no longer grows eventually forms \citep{RajaratnamBeltaos1977}. More recent work by \citet{Gorman2023} found that the depth, $d$ of a thruster-formed crater initially grows linearly with thrusting duration, $t$, such that $d \propto t$. After this initial linear phase, $d$ begins to grow sub-linearly: $d \propto t^{1/3}$. Interestingly, this sub-linear growth of $d$ is consistent with the $\pi$-scaling framework, if we consider that the total interacting mass, $m_g = \dot m_g t$. From the definition of $\pi_{V,g}$, $V_c \propto m$. Since $d \propto V^{1/3}$, then $d \propto \left(\dot m_g t\right)^{1/3}$. We found that when adopting this relationship between the mass flux and total interacting mass, Eq.\ref{eq:pi_gas_equation_general} can predict the outcomes of literature gas excavation results that have $t \sim 0.5$ s \citep{MasonNordmeyer1969}.

\textbf{3) Gas-crater ejecta-scaling relationships.} Determining whether a thruster maneuver can produce particles with velocities low enough to remain gravitationally bound to an asteroid requires measurement of the ejecta velocity distribution. The high-speed camera data acquired in our study can provide a starting point for measuring particle trajectories, ejection angles, and speeds. However, they are limited in their ability to capture the full three-dimensional distribution, as out experiments were designed for measuring crater volume. Applying ejecta velocity distribution analysis to the existing experiments and to future tests conducted under conditions closer to an operational asteroid scenario will enable the development of gas-crater ejecta-scaling relationships analogous to those used for impact cratering \citep{Holsapple2012}. Such relationships would connect maneuver and target properties to the amount and velocity distribution of excavated material available for spacecraft tracking.

\textbf{4) A seismic source function for gas-based craters.} Seismometers proposed for asteroid interior exploration \citep[e.g.,][]{Ballouz+2024,dellagiustina_seismology_2024,murdoch_probing_2017,murdoch_seismic_20252025LPICo3083.2036M}  require a natural or artificial source of seismic energy. Natural meteoroid impacts may be rare relative to the operational lifetime of a deployed seismometer, whereas spacecraft thrusters provide a controllable source already available on a rendezvous spacecraft. Future work should characterize the source-time function, frequency content, repeatability, and efficiency with which gas thrusting couples energy into the surface. These measurements will determine whether thruster-driven excavation can provide a sufficiently strong and interpretable artificial seismic source for characterizing asteroid interiors.

%% file: Tables/Discussion_Summary_Table.tex
\begin{deluxetable*}{lcccc}
\tablecaption{Summary of target, gas, and $\pi$-scaling
parameters for the cases considered in this study.
\label{tab:pi_scaling_summary}}

\tablehead{
\colhead{Parameter} &
\colhead{1$g$} &
\colhead{$\sim10^{-4}g$} &
\colhead{MN1969 Small} &
\colhead{MN1969 Large}
}

\startdata
\cutinhead{Target properties}
Average grain size, $\bar D$ (mm) &
$4.4$ & $0.53$ & $0.26$ & $0.26$ \\
Bulk density, $\rho$ (kg m$^{-3}$) &
$1652\pm2$ & $1320$ & $1100$ & $1100$ \\
Gravity, $g$ (m s$^{-2}$) &
$9.81$ & $\sim3\times10^{-3}$ & $9.81$ & $9.81$ \\
Strength, $Y$ (kPa) &
$0$ & $0$ & $3.4$ & $3.4$ \\
\cutinhead{Source properties}
Nozzle height, $h$ (mm) &
$-10$--$100$ & $70$ &
$68.58$--$137.16$ & $139.70$--$304.80$ \\
Nozzle diameter, $a$ (mm) &
$9.5$ & $0.5$ & $34.5$ & $138.7$ \\
Gas density, $\delta$ (kg m$^{-3}$) &
$0.47$--$3.3$ & $1.73$ &
$(0.753\text{--}2.26)\times10^{-3}$ &
$(2.73\text{--}3.89)\times10^{-3}$ \\
Gas speed, $U$ (m s$^{-1}$) &
$480$ & $321$ &
$2.75\times10^{3}$ & $2.75\times10^{3}$ \\
\cutinhead{Scaling constants}
$\mu$ &
$0.55$ & $0.55$ &
$0.55$ & $0.55$ \\
$K_1$ &
$0.77$ & $0.77$ &
$0.77$ & $0.77$ \\
$K_2$ &
$0$ & $0$ & $0.45$ & $1.98$ \\
\enddata

\tablecomments{
Ranges encompass the individual experiments included in each grouped analysis. The gas densities and gas speeds for the MN1969 experiments \citep{MasonNordmeyer1969} were reconstructed using the assumptions described in the text rather than taken from reported measurements.
}
\end{deluxetable*}

%% file: _05_acknowledgements.tex
\begin{acknowledgments}
This work was supported by the Johns Hopkins Applied Physics Laboratory and a NASA Yearly Opportunities for Research in Planetary Defense grant, award 80NSSC25K7434. C.S. acknowledges support from CNES. ChatGPT-5.6 Sol was used to convert tables and equations to the \LaTeX format.
\end{acknowledgments}

\begin{contribution}
\end{contribution}

%% file: _07_appendix.tex
\section{Appendix}
\label{sec:appendix}

\subsection{Calculating Experimental Gas Properties at the Nozzle Exit}
\label{sec:appendix_1d_gas_model}

\begin{figure}
    \centering
    \includegraphics[width=\columnwidth]{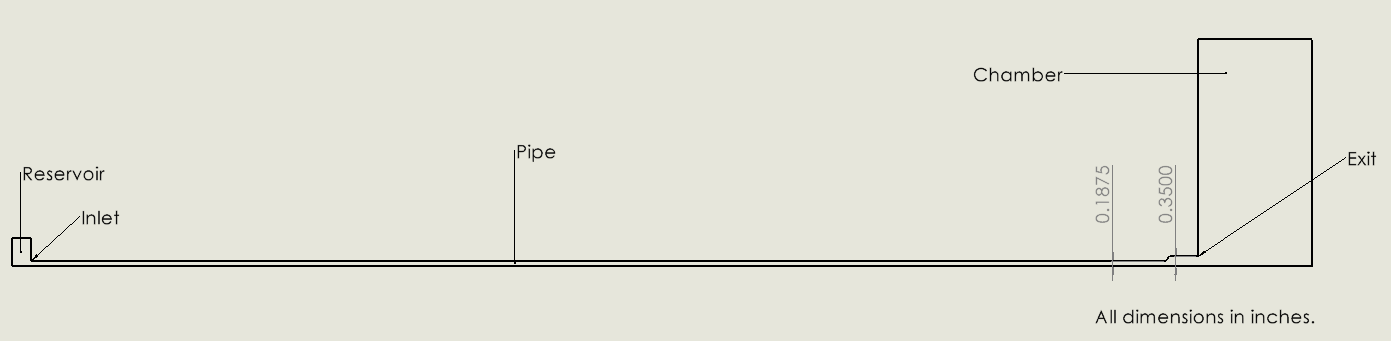}
    \caption{Labeled Drawing of ANSYS Fluent simulation domain. The bottom edge is the axis of symmetry.}
    \label{fig:fluent_domain}
\end{figure}

We calculate the down-stream properties of the gas using  computational fluid dynamics (CFD) simulations with \citet{ansys_fluent_2023}. The simulations consider the release of gas from the reservoir, its expansion through the barrel,  its acceleration at the barrel's ``nozzle'', and its venting into the vacuum chamber. The geometry and initial conditions were selected to match the corresponding laboratory experiments. 
Specifically, the modeled system consisted of a reservoir with a volume of $3.27741\times10^{-5}$~m$^3$, connected to a $1$-m-long pipe with an inner diameter of $3/8$~in with an expansion at the end that increased the diameter to $0.7$~in. The computational domain was constructed and meshed with the geometry shown in Fig.~\ref{fig:fluent_domain}, with the pressurized reservoir discharging through the pipe and expanding into a downstream chamber. The dimensions of the chamber were selected primarily to provide a computational domain surrounding the pipe exit.
A two-dimensional axisymmetric domain was used to reduce computational cost and allow for finer meshing. The lower boundary shown in Fig.~\ref{fig:fluent_domain} was defined as the axis of symmetry, such that the two-dimensional solution represents the corresponding three-dimensional geometry upon revolution about this axis.

\begin{figure}
    \centering
    \includegraphics[width=\columnwidth]{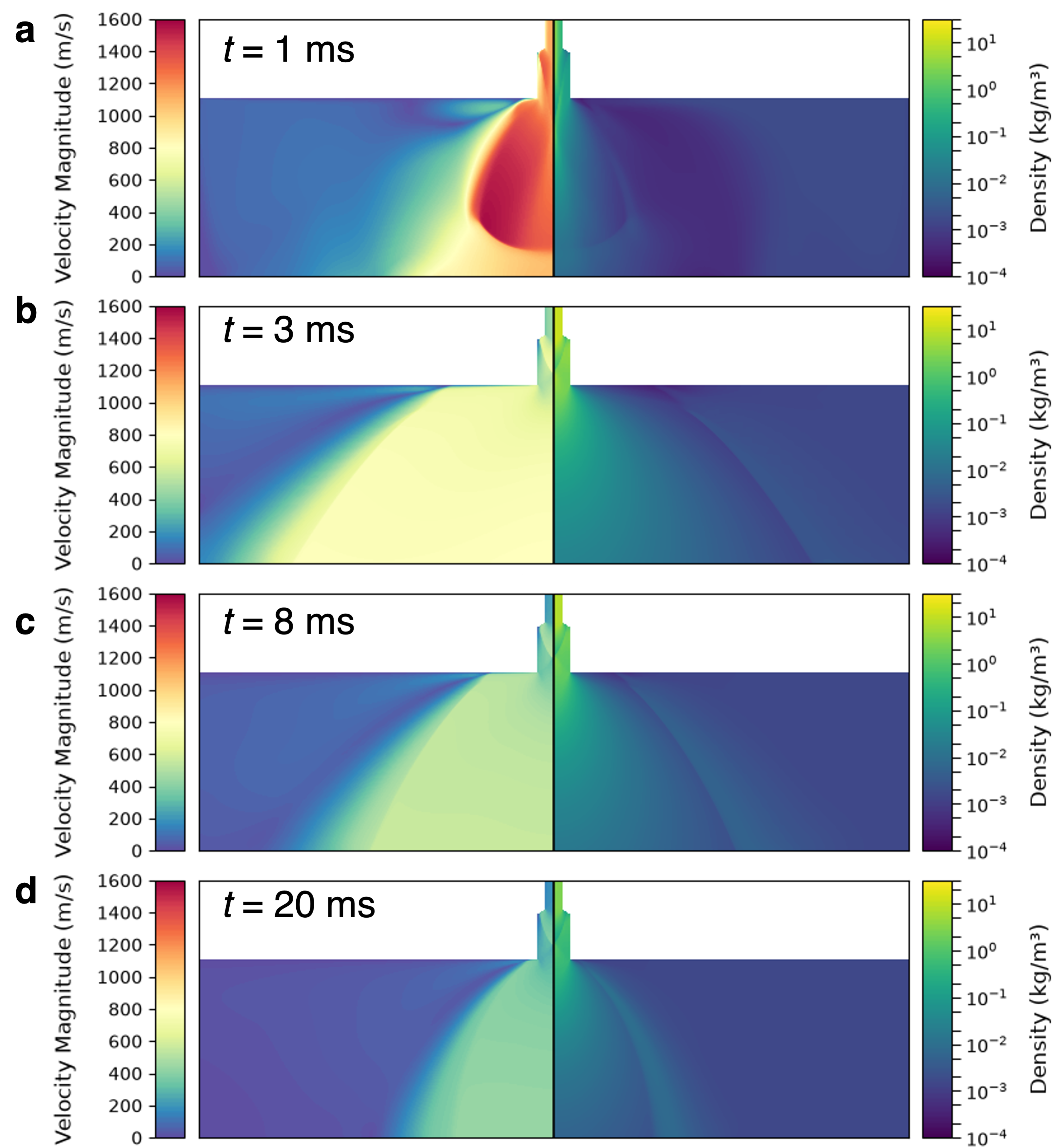}
    \caption{Results of Fluent simulations with initial reservoir pressure of 1000 psi over time. The left panels show gas velocity, $U$, and the right panels show gas density, $\delta$. The gas reaches super-sonic speeds in the initial few milliseconds and expand at a wide angle before narrowing. Calculation results at time, $t = $ \textbf{a}, 1 ms, \textbf{b} 3 ms, \textbf{c}, 8 ms, and \textbf{d}, 20 ms.}
    \label{fig:fluent_unsteady}
\end{figure}

\begin{figure}
    \centering
    \includegraphics[width=0.5\columnwidth]{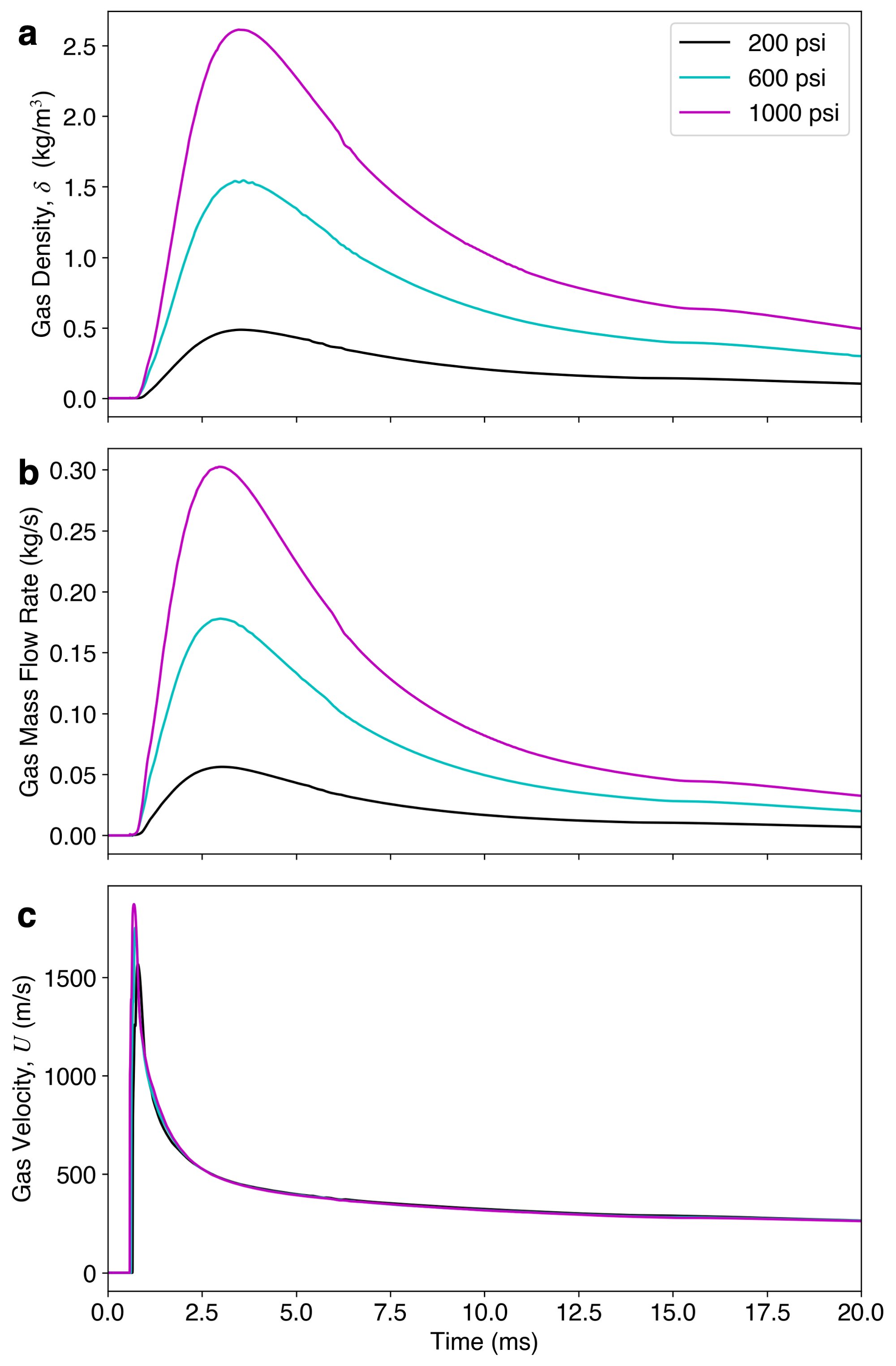}
    \caption{Fluent simulation results of gas properties at the nozzle exit as a function of time for cases with nominal experimental conditions and reservoir initial static pressures of 200 psi (black curves), 600 psi (cyan curves), and 1000 psi (magenta curves). \textbf{a,} $\delta$ at nozzle exit. \textbf{b,} gas mass flow rate at nozzle exit. \textbf{c}, gas velocity at nozzle exit.}
    \label{fig:fluent_results}
\end{figure}

The chamber boundaries were defined as pressure outlets with a static pressure of $1$~torr and a temperature of $293$~K, reflecting the nominal conditions of the laboratory vacuum chamber. All remaining solid boundaries were modeled as no-slip walls. The computational domain was initially filled with nitrogen at the chamber conditions. Following initialization, the reservoir volume was patched to a set pressure depending on the experiment with zero velocity, representing a pressurized reservoir containing initially stagnant gas. This configuration approximates the experimental procedure in which an evacuated pipe and chamber are connected to a pressurized reservoir containing gas at rest and, when prompted, a valve is opened to release the gas. We conduct unsteady flow simulations. At the start of the transient simulation, the valve is opened and the pressure differential drives the nitrogen from the reservoir through the pipe and into the evacuated chamber (Fig.~\ref{fig:fluent_unsteady}). 
The transient simulation was run for a duration of $20$~ms, which was sufficient to capture the peaks of the relevant flow properties at the reservoir, pipe inlet, and pipe exit, including mass flow rate, velocity, temperature, pressure, and density. An area-weighted average is performed over the cells that comprise the nozzle respective point of interest and these properties are shown for three of the simulations in Fig.~\ref{fig:fluent_results}.

